\documentclass[aps,pra,reprint,amsmath,amssymb,superscriptaddress,longbibliography]{revtex4-2}

\usepackage{graphicx}
\usepackage[colorlinks=true,urlcolor=blue,citecolor=blue,linkcolor=blue,bookmarks=false,pdfstartview={FitH}]{hyperref}

\usepackage{color,soul}

\begin{document}

\title{Crystal-field excitations and vibronic modes in triangular-lattice spin-liquid candidate TbInO$_3$}

\author{Mai Ye}
\email{mye@physics.rutgers.edu}
\affiliation{Department of Physics and Astronomy, Rutgers University, Piscataway, NJ 08854, USA}
\author{Xianghan Xu}
\affiliation{Department of Physics and Astronomy, Rutgers University, Piscataway, NJ 08854, USA}
\author{Xiangyue Wang}
\affiliation{Department of Physics and Astronomy, Rutgers University, Piscataway, NJ 08854, USA}
\author{Jaewook Kim}
\affiliation{Department of Physics and Astronomy, Rutgers University, Piscataway, NJ 08854, USA}
\author{Sang-Wook Cheong}
\affiliation{Department of Physics and Astronomy, Rutgers University, Piscataway, NJ 08854, USA}
\author{Girsh Blumberg}
\email{girsh@physics.rutgers.edu}
\affiliation{Department of Physics and Astronomy, Rutgers University, Piscataway, NJ 08854, USA}
\affiliation{Laboratory of Chemical Physics, National Institute of Chemical Physics and Biophysics, 12618 Tallinn, Estonia}

\date{\today}

\begin{abstract}
We study the ground state properties, the electronic excitations and lattice dynamics in spin-liquid candidate TbInO$_3$. 
By employing polarization resolved Raman spectroscopy we define the inter- and intra-multiplet excitations, and establish the low-energy crystal-field (CF) level scheme. 
In particular, we demonstrate that the ground state of the Tb$^{3+}$ ions is a non-Kramers doublet, and relate the enhanced linewidth of the CF modes to the magnetic fluctuations near the spin-liquid ground state.
We identify the 38 allowed Raman-active phonon modes at low temperature. 
Moreover, we observe hybrid vibronic excitations involving coupled CF and low-lying phonon modes, suggesting strong spin-lattice dynamics.
We develop a model for vibronic states and obtain the parameters of the bare responses and coupling strength. 
We further demonstrate that the obtained CF level scheme is consistent with specific heat data.
\end{abstract}

\maketitle

\section{Introduction\label{sec:Intro}}

Interplay between electron correlation and spin-orbit coupling (SOC) gives rise to a variety of emergent quantum phases and transitions~\cite{Takayama2021}. 
Especially, because of the compactness of $f$ orbits and the heaviness of $f$ elements, $f$-electron systems are in both the strong Mott regime and the strong SOC regime~\cite{Witczak-Krempa2014}. 
For these materials, exotic spin-liquid (SL) and multipolar-ordered ground states were predicted~\cite{Suzuki2018,Takagi2019}. 
The multipolar interactions and the resulting ordering phenomena have also been studied~\cite{Santini2009,Kuramoto2009}. However, rare-earth-based SL systems still remain to be explored, especially from the experimental side; many of the candidate materials are not widely accepted as hosting SL ground state due to absence of convincing experimental evidence.

\begin{figure}
\includegraphics[width=0.48\textwidth]{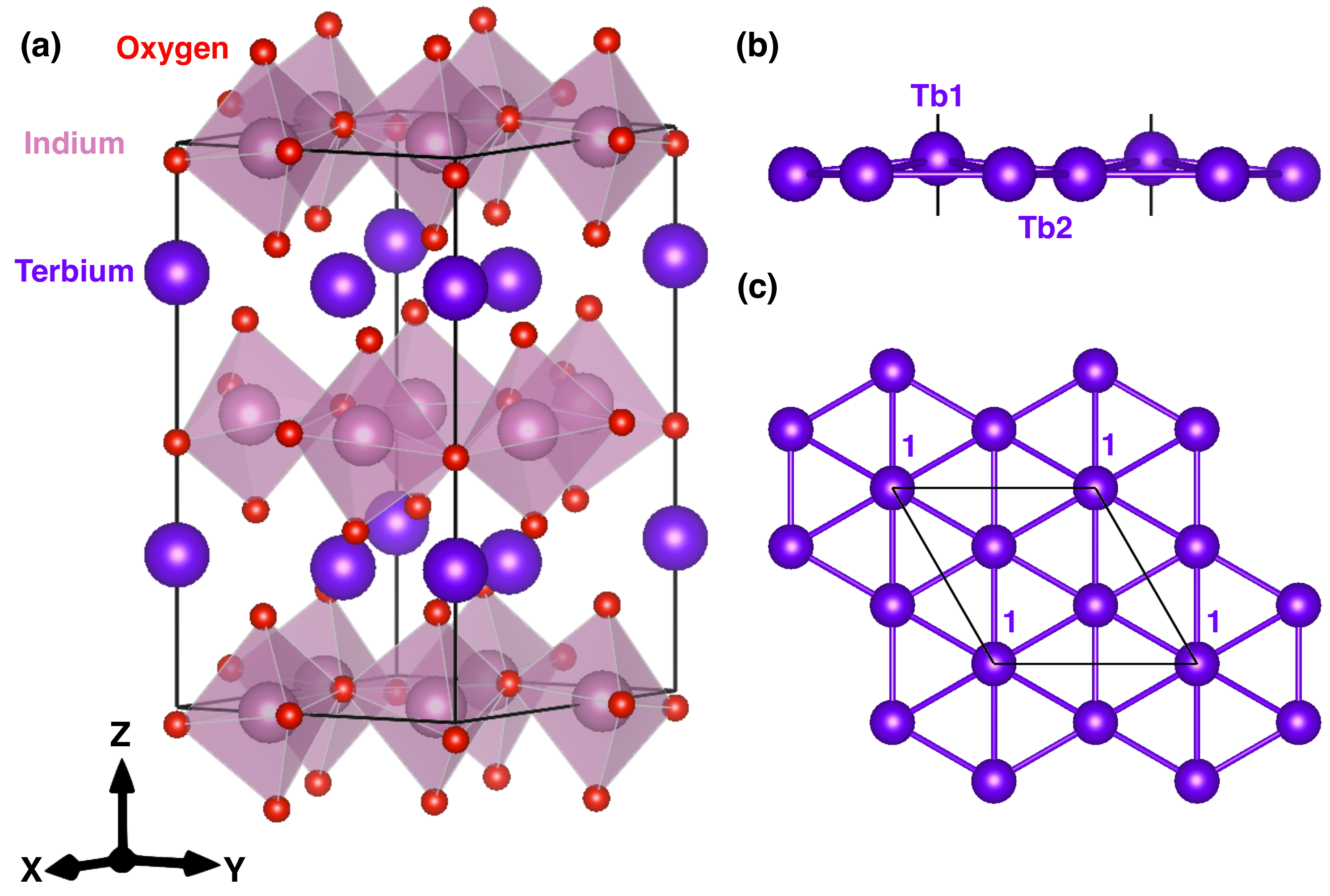}
\caption{\label{fig:S}
Crystal structure of ferroelectric TbInO$_3$. 
(a) The black frame indicates the unit cell. 
(b) The Tb layer viewed from X direction. The Tb2 sites form a hexagonal lattice while the Tb1 sites slightly buckle out of the plane. (c) The Tb layer viewed from Z direction. The Tb1 sites at the hexagonal centers are labeled by "1".}
\end{figure}

It has been recently proposed that ferroelectric insulator TbInO$_3$ could harbor a 2D spin-liquid ground state~\cite{Gaulin2019,Kiryukhin2019,Cheong2019}.
In the ferroelectric phase this material has a hexagonal structure~\cite{Pistorius1976} (space group $P6_3cm$, No.\,185; point group C$_{6v}$), as shown in Fig.\,\ref{fig:S}(a). 
The magnetic Tb$^{3+}$ ions form a slightly distorted triangular lattice, separated by non-magnetic layers of corner-sharing [InO$_5$]$^{7-}$ polyhedra. The Tb$^{3+}$ ions possess two different site symmetries: Tb2 sites form a hexagonal plane, while Tb1 sites reside at the hexagonal centers and buckle slightly out of the hexagonal plane [Fig.~\ref{fig:S}(b-c)]. Hence the site symmetry is C$_{3v}$ for Tb1 sites and C$_{3}$ for Tb2 sites. The distortion from ideal triangular lattice is weak because the buckling of Tb1 is only 0.38\,$\AA$, around one tenth of the in-plane Tb2-Tb2 distance (3.65\,$\AA$)~\cite{Gaulin2019}. The fluctuating magnetic moment is shown to be confined to the triangular-lattice plane~\cite{Kiryukhin2019}. Magnetic susceptibility of polycrystalline samples obeys Curie-Weiss law above 10\,K, with the Weiss temperature being -17.2\,K; however, no magnetic ordering or spin freezing occurs down to 0.1\,K~\cite{Gaulin2019}, indicating strong magnetic frustration.

The classification of the proposed spin-liquid state in TbInO$_3$, though, remains unclear. The inelastic neutron-scattering (INS) study by L.\,Clark et al. suggests a singlet crystal-field (CF) ground state for Tb1, and a doublet for Tb2 ions~\cite{Gaulin2019}. 
At low temperature, then, Tb1 ions become nonmagnetic, and a honeycomb lattice of magnetic Tb2 ions emerges. In this scenario a Kitaev Z$_2$-symmetry SL is likely realized~\cite{Kitaev2006}. 
This CF level scheme was supported by analysis of the low-temperature specific heat~\cite{Cheong2019}. 
On the contrary, the observations of another INS study by M.G.\,Kim et al. are consistent with a SL state based on triangular lattice~\cite{Kiryukhin2019}. Such a scenario implies that the CF ground states of Tb1 and Tb2 are the same, and the SL state might have a continuous symmetry. 
Interestingly, if the CF ground states of Tb1 and Tb2 are both non-Kramers doublets, on a triangular lattice spin-orbit-entangled multipolar phases are predicted to emerge~\cite{Liu2018}. 
It is therefore vital to identify the ground state of the Tb$^{3+}$ ions in this system, because this is the principal step to construct the spin Hamiltonian and to elaborate on emergent low-energy physics.

Moreover, in rare-earth compounds vibronic modes, namely coupled vibrational and CF electronic excitations, have been observed~\cite{Cooper2019,Cardona1991,Gaulin2018}.
Thus, the lattice may play a pivotal role in defining the electronic ground state.  
Therefore, we explore electron-phonon interactions and such hybrid modes in TbInO$_3$. 

In this work, we present spectroscopic study of the electronic and phononic excitations in TbInO$_3$.
We measure the inter- and intra-multiplet excitations, and establish the CF level scheme within the lowest-energy multiplet. 
We find that the ground state of the Tb$^{3+}$ ions is a non-Kramers doublet, and show that the derived CF level scheme is consistent with the entropy data. 
We determine the energy of the 38 allowed Raman-active phonon modes at low temperature.
In addition, we observe hybrid vibronic excitations resulting from strong coupling between the CF and phonon modes. The coupling originates from the modulation of the electron-cloud distribution of the CF states by lattice vibration.

The rest of this paper is organized as follows. In Sec.~\ref{sec:Exp} we describe the sample preparation and experimental setup. In Sec.~\ref{sec:Inter} we present the inter-multiplet excitations. In Sec.~\ref{sec:CF} we discuss the CF excitations within the lowest-energy multiplet, and establish the CF level scheme. In Sec.~\ref{sec:P} we show the phonon spectra at low temperature and identify the Raman-active phonon modes. In Sec.~\ref{sec:Vib} we examine the hybrid vibronic spectral features, and explain how the coupling between the CF and phonon modes leads to such vibronic features. In Sec.~\ref{sec:SH} we analyze the specific heat and entropy data, demonstrating their consistency with the CF level scheme. In Sec.~\ref{sec:Con} we provide a summary of the observations and their implications. 
The relevant mathematical formalisms are provided in appendices: 
classification of the CF states in each multiplet of the $^7F$ term in Appendix~\ref{sec:AM}; 
classification of the $\Gamma$-point phonons in Appendix~\ref{sec:AP}; 
expressions relevant to the coupling between one phonon and one CF mode in Appendix~\ref{sec:AVibSimple}; 
the fitting model for the vibronic features in Appendix~\ref{sec:AVib}; the specific-heat model in Appendix~\ref{sec:ASH}. 

\section{Experimental\label{sec:Exp}}

Single crystals of TbInO$_3$ were prepared using laser-floating-zone method, and characterized by Laue diffraction to confirm single phase~\cite{Cheong2019}.
Two samples were used for Raman study: one cleaved to expose its (001) crystallographic plane and the other polished to expose its (010) plane.
The samples were then examined under a Nomarski microscope to find a strain-free area.
Raman-scattering measurements were performed in a quasi-back-scattering geometry from the samples mounted in a continuous helium gas flow cryostat.

We used a custom fast f/4 high resolution 500/500/660\,mm focal lengths triple-grating spectrometer for data acquisition. 
All the data were corrected for the spectral response of the spectrometer.

For acquisition of the low-frequency Raman response, we used 1800\,mm$^{-1}$ master holographic gratings; the 647 and 676\,nm lines from a Kr$^+$ ion laser were used for excitation: 647\,nm laser line combined with 100\,$\mu$m slit width provides 0.19\,meV spectral resolution; 676\,nm laser line combined with 25\,$\mu$m slit width provides 0.05\,meV spectral resolution. 
For the high-frequency Raman response, we used 150\,mm$^{-1}$ ruled gratings; the 476 and 568\,nm lines from the same Kr$^+$ ion laser were used for excitation: 476\,nm laser line combined with 100\,$\mu$m slit width provides around 4\,meV spectral resolution; 568\,nm laser line combined with 100\,$\mu$m slit width provides around 3\,meV spectral resolution.

For polarization optics, a Glan-Taylor polarizing prism (Melles Griot) with a better than 10$^{-5}$ extinction ratio to clean the laser excitation beam and a broad-band 50\,mm polarizing cube (Karl Lambrecht Corporation) with an extinction ratio better than 1:500 for the analyzer was used. To perform measurements with circularly-polarized light, we use a Berek polarization compensator (New Focus) after the polarizing prism to convert the incoming linearly-polarized light into circularly-polarized light for excitation, and a broad-band 50\,mm-diameter quarter wave retarder (Melles Griot) before the polarizing cube to convert the outcoming circularly-polarized light into linearly-polarized light for the analyzer. Experiments involving circular polarization geometry are performed from the (001) crystallographic plane.

Incident light was focused to a 50$\times$100\,$\mu$m$^{2}$ spot. For the spectra at 5\,K, 2\,mW laser power was used; for other data, 5\,mW laser power of was used. 
The reported temperature values were corrected for laser heating: 1\,K$\slash$\,mW laser heating rate was assumed.

The measured secondary-emission intensity $I(\omega,T)$ is related to the Raman response $\chi''(\omega,T)$ by $I(\omega,T)=[1+n(\omega,T)]\chi''(\omega,T)$, where $n$ is the Bose factor, $\omega$ is energy, and $T$ is temperature.

\begin{figure}[t]
\includegraphics[width=0.46\textwidth]{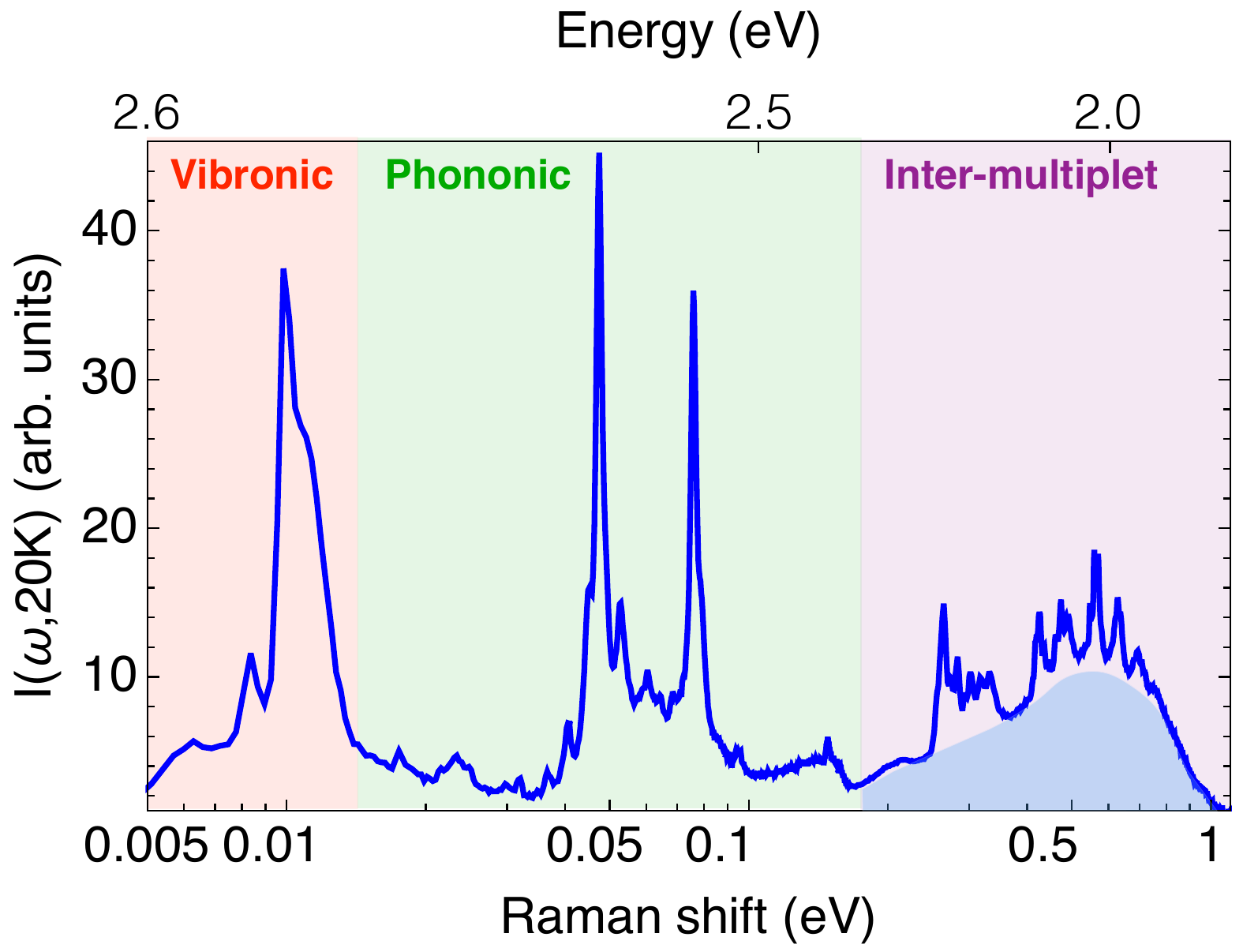}
\caption{\label{fig:OV} 
An overview of the low-temperature secondary-emission intensity $I$ measured in XY geometry at 20\,K with 476\,nm excitation in semi-log scale. 
The top scale is the absolute energy of the secondary-emission photons; 
the bottom scale shows the Raman shift: the energy loss, or the laser-photon energy minus the scattered-photon energy. 
The region below 15\,meV features coexistence of CF and phononic excitations, and the resulting hybrid vibronic modes. The region between 15 and 80\,meV is dominated by phonon modes. The region above 80\,meV contains the inter-multiplet excitations on top of a broad photo-luminescence continuum (shaded in blue).}
\end{figure}
In Fig.\,\ref{fig:OV} we show an overview spectrum to illustrate the relevant energy scales. Below 15\,meV, energy proximity between the CF state of the lowest-energy multiplets and the phonon modes leads to vibronic spectral features. From 15\,meV to 80\,meV, phononic features dominate. Above 80\,meV, the spectrum is composed of the inter-multiplet excitations on top of a broad photo-luminescence continuum.

\section{Inter-multiplet excitations\label{sec:Inter}}

\begin{figure}
\includegraphics[width=0.46\textwidth]{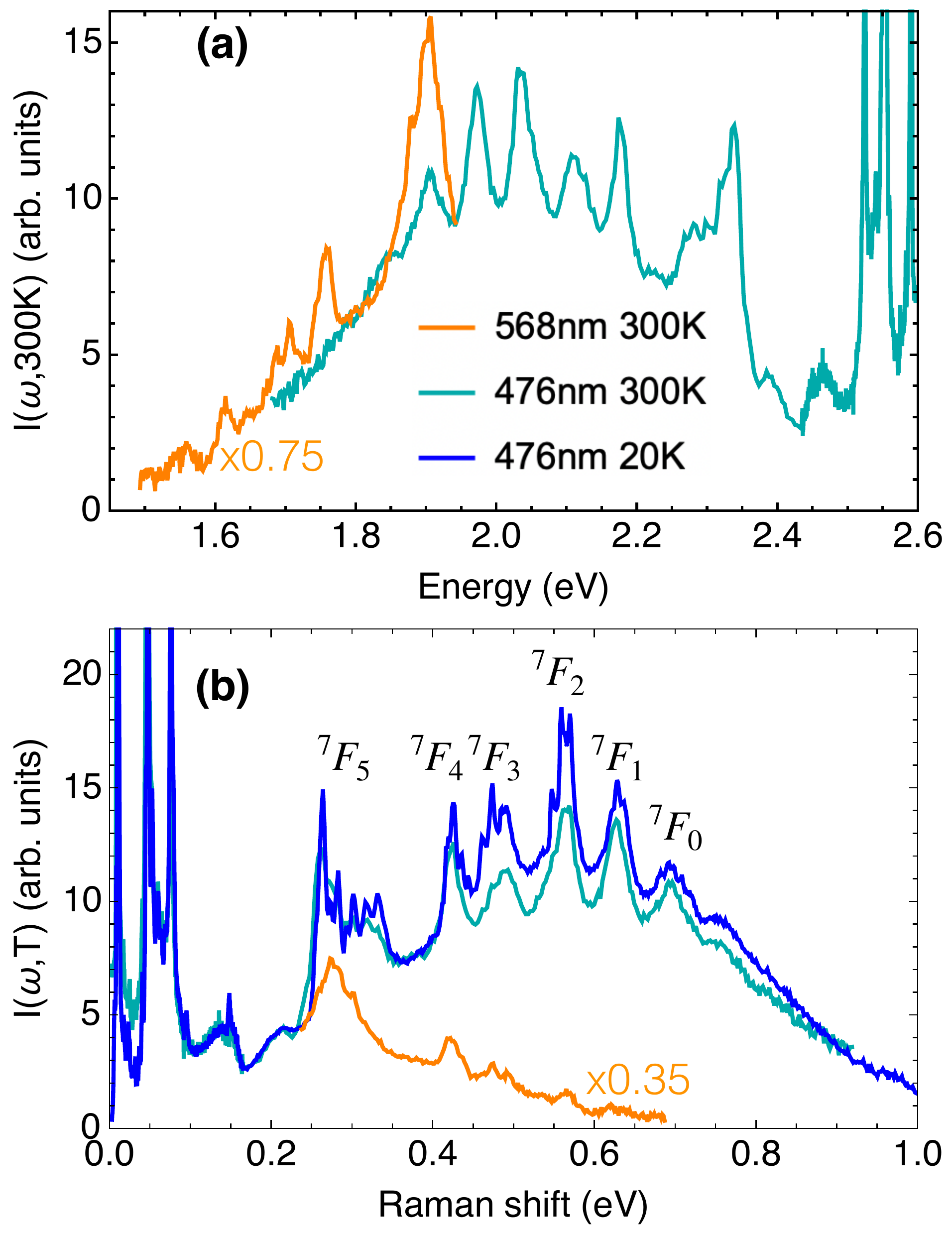}
\caption{\label{fig:M}The inter-multiplet excitations within the $^7F$ term of Tb$^{3+}$ ions in TbInO$_3$. Secondary-emission intensity $I$ in XY scattering geometry is measured. (a) Secondary-emission intensity measured at 300\,K as a function of outcoming photon energy. (b) Secondary-emission intensity measured at 300 and 20\,K as a function of Raman shift.}
\end{figure}

In this section we present the inter-multiplet excitations within the $^7F$ term of Tb$^{3+}$ ions.
The electronic configuration of Tb$^{3+}$ ion is 4$f^8$. 
According to Hund's rules, the $^7F$ term has the lowest energy. 
In Fig.\,\ref{fig:M} we show the inter-multiplet excitations within the $^7F$ term. They are on top of a broad photo-luminescence continuum centered at 2.05\,eV [Fig.\,\ref{fig:M}(a)]. 
This luminescence features could be attributed to the dipole-allowed ${^5D}\,\rightarrow\,{^7F}$ transition~\cite{Mansouri2018,Couwenberg1998,Carnall1989}.

The ground-state multiplet should have the largest possible total angular momentum, namely $J=6$. The other multiplets of the $^7F$ term, with total angular momentum from $J=5$ to $J=0$ in integer step, have increasing energy in order: 0.30, 0.43, 0.48, 0.56, 0.63 and 0.70\,eV, respectively [Fig.\,\ref{fig:M}(b)]. The energies of the multiplets measured in this study are consistent with those measured in TbMnO$_3$~\cite{Mansouri2018}, TbAl$_3$(BO$_3$)$_4$~\cite{Couwenberg1998}, and Tb-doped LaF$_3$~\cite{Carnall1989}.

These multiplets are further split by the CF potential, resulting in sharp spectral features clustered within a particular energy range. 
Fine intra-multiplet CF structures are revealed at low temperature [Fig.\,\ref{fig:M}(b)]. 
Because the degeneracy of each multiplet is $2J+1$, the number of CF levels within each multiplet decreases with $J$ [Appendix~\ref{sec:AM}].

\section{Crystal-field excitations\label{sec:CF}}

In this section we discuss the intra-multiplet excitations, between the CF levels, within the $^7F_6$ ground multiplet. 
The general expression for a CF potential of C$_{3v}$ site symmetry can be written as~\cite{Hutchings1964}
\begin{equation}
H_{CF}=B_2^0\hat{O}_2^0+B_4^0\hat{O}_4^0+B_4^3\hat{O}_4^3+B_6^0\hat{O}_6^0+B_6^3\hat{O}_6^3+B_6^6\hat{O}_6^6~,
\label{eq:HCF}
\end{equation}
where $\hat{O}$'s are Stevens operators~\cite{Stevens1952} and the $B$'s are the CF coefficients. The C$_{3v}$-symmetry CF potential splits the 13-fold degenerate $^7F_6$ multiplet, resulting in 3\,A$_{1}$ singlets, 2\,A$_{2}$ singlets and 4\,E doublets. 
The CF wavefunctions can be expressed in the $|J,m_j\rangle$ bases; by group theory we can identify the basis functions of the CF eigenfunctions of each symmetry. 
These basis functions are given in Table\,\ref{tab:F2}. 
The A$_{1}$ and A$_{2}$ singlets are non-magnetic, while the E doublets allow a finite magnetic dipole moment.
On the site symmetry reduction from C$_{3v}$ to C$_{3}$, the A$_{1}$ and A$_{2}$ states of C$_{3v}$ group merge into the A states of C$_{3}$. 
Therefore, the C$_{3}$-symmetry CF potential splits the $^7F_6$ multiplet into 5\,A singlets and 4\,E doublets.
Due to the mirror symmetry breaking, both the A singlets and E doublets of C$_{3}$ group could carry magnetic moment.

\begin{table}
\caption{\label{tab:F2}The basis functions of the A$_{1}$-, A$_{2}$- and E-symmetry CF eigenfunctions, respectively. These basis functions are expressed as combinations of $|m_j\rangle$ with J=6.}
\begin{ruledtabular}
\begin{tabular}{ccc}
A$_{1}$&A$_{2}$&E\\
\hline
$\frac{1}{\sqrt{2}}(|+3\rangle-|-3\rangle)$&$\frac{1}{\sqrt{2}}(|+3\rangle+|-3\rangle)$&$\frac{1}{\sqrt{2}}(|+1\rangle\pm|-1\rangle)$\\
$\frac{1}{\sqrt{2}}(|+6\rangle+|-6\rangle)$&$\frac{1}{\sqrt{2}}(|+6\rangle-|-6\rangle)$&$\frac{1}{\sqrt{2}}(|+2\rangle\pm|-2\rangle)$\\
$|0\rangle$&&$\frac{1}{\sqrt{2}}(|+4\rangle\pm|-4\rangle)$\\
&&$\frac{1}{\sqrt{2}}(|+5\rangle\pm|-5\rangle)$\\
\end{tabular}
\end{ruledtabular}
\end{table}

\begin{figure}[t]
\includegraphics[width=0.46\textwidth]{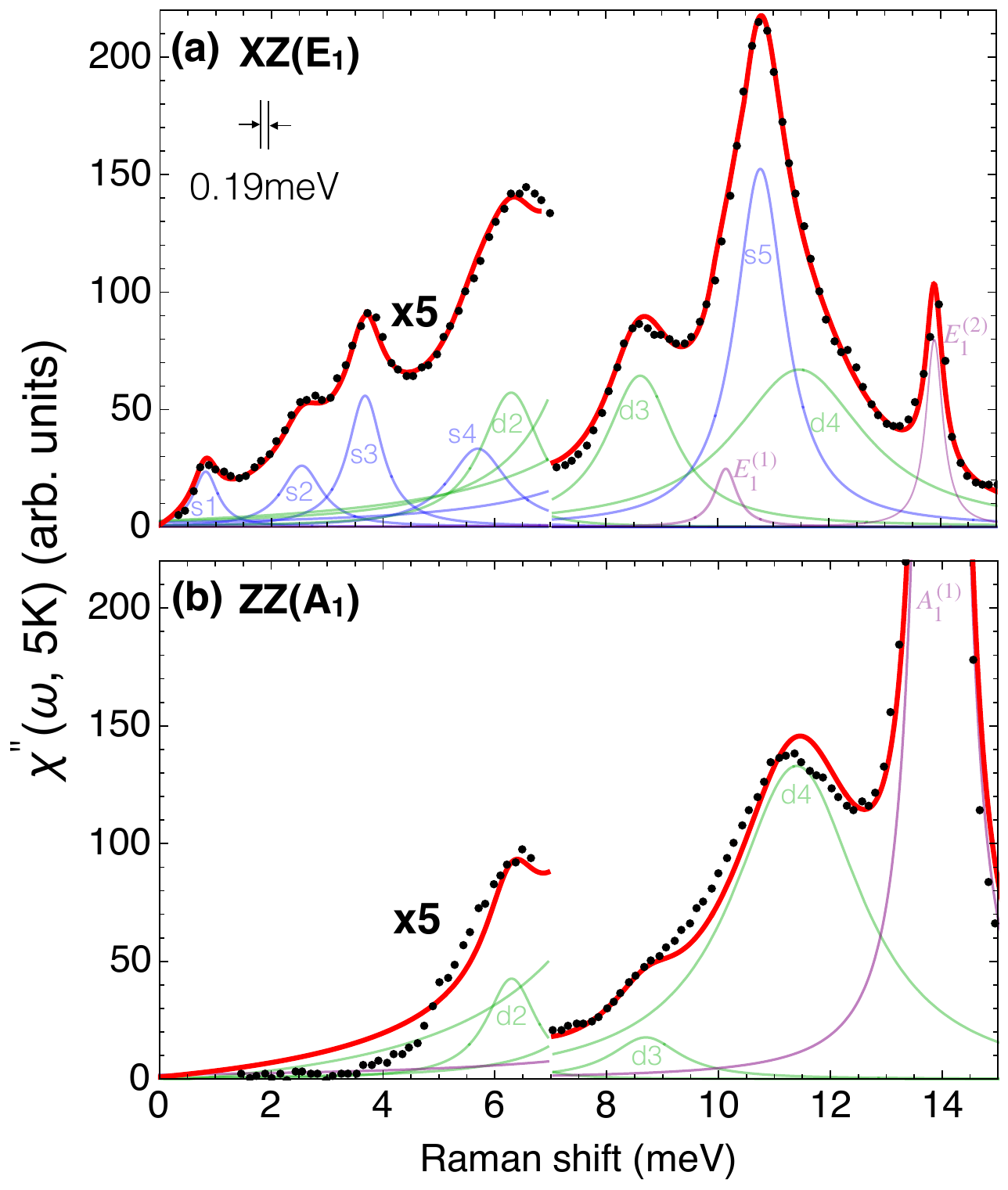}
\caption{\label{fig:CF1}The low-energy crystal-field (CF) transitions in TbInO$_3$, measured at 5\,K with 647\,nm excitation in (a) XZ and (b) ZZ scattering geometries.
The Raman data are represented by black dots. 
The red lines represent the fits by Lorentzian lineshapes. 
The blue lines show doublet-to-singlet CF oscillators labeled by the final singlet state, and the green lines show doublet-to-doublet CF transitions; 
the phonon modes, labeled by their symmetry, are shown in purple. 
The Raman data and fitting curves below 7\,meV are multiplied by a factor of 5 for clarity.}
\end{figure}

In Fig.\,\ref{fig:CF1} we present the high-resolution spectra of the CF excitations within the $^7F_6$ multiplet. The relationship between the scattering geometries and the symmetry channels is given in Table~\ref{table:Exp1}: the singlet-to-singlet transitions (A symmetry) appear in ZZ geometry; the doublet-to-singlet transitions (E symmetry) appear in XZ and RL geometries; the doublet-to-doublet transitions (A and E symmetries) appear in all these three geometries. 
We note that if Tb1 and Tb2 have different CF level scheme, at least 6 CF modes should appear in ZZ geometry,  however, only 3 CF modes are resolved [Fig.\,\ref{fig:CF1}(b)]. 
Hence both sites have essentially the same CF level scheme, consistent with the weakness of ferroelectric distortion~\footnote{The same CF level scheme for both Tb1 and Tb2 is further confirmed by the observation of 8 crystal-field (CF) modes in XZ geometry [Fig.~\ref{fig:CF1}(a)], and the temperature dependence of CF entropy [Fig.~\ref{fig:SH}(b)]}.

\begin{table}[b]
\caption{\label{table:Exp1}
Relationship between the scattering geometries and the symmetry channels. 
Each geometry is represented by E$_{i}$E$_{s}$, where E$_{i}$ and E$_{s}$ are the polarizations of incident and scattered light; X, Y, and Z are the [100], [010], and [001] crystallographic directions; R and L represent right and left circular polarizations. 
The symmetry of the phonon modes is classified by the irreducible representations of C$_{6v}$ group; the symmetry of the CF transitions at Tb1 and Tb2 sites is classified by those of C$_{3v}$ group and C$_{3}$ group, respectively.}
\begin{ruledtabular}
\begin{tabular}{lllll}
Scattering geometry      & RL      & XZ      & ZZ      \\
\hline
Phonon modes (C$_{6v}$)  & E$_{2}$ & E$_{1}$ & A$_{1}$ \\
CF transition (C$_{3v}$) & E       & E       & A$_{1}$ \\
CF transition (C$_{3}$)  & E       & E       & A       \\ 
\end{tabular}
\end{ruledtabular}
\end{table}

First, we fit the XZ spectrum with Lorentzian functions to determine the energy and linewidth of the CF excitations [Fig.~\ref{fig:CF1}(a)]. 
Because we find eight CF modes in XZ spectrum, the ground state must be a doublet: otherwise there would be only four CF modes in XZ spectrum.

Next, we turn to the ZZ spectrum. 
Because the CF transitions at 0.8, 2.5 and 3.7\,meV are absent in the ZZ spectrum [Fig.~\ref{fig:CF1}(b)], we assign the states as singlets (the doublet-to-singlet transitions should not appear in ZZ geometry). 
For the remaining CF transitions, there are two in 5-7\,meV range and three in 8-12\,meV range. 
We find that to fit the A$_{1}$-symmetry spectrum, one mode in 5-7\,meV interval and two modes in 8-12\,meV interval are required. 
These three modes must be doublets, and there are only three doublets to be assigned. 
Hence, the two modes in 5-7\,meV interval must be a singlet and a doublet. 
Among them, we assign the 6.3\,meV mode as a doublet to better fit the A$_{1}$-symmetry spectrum~\footnote{The energy and width of a CF mode are intrinsic properties of the system which is same for all symmetry channels. This requirement puts a constraint on the fitting process.}. 
For the same reason, we assign the 8.7 and 11.5\,meV modes in 8-12\,meV range as doublets. 
The CF energies and transition widths are given in Table\,\ref{table:CF}.

\begin{table}[b]
\caption{\label{table:CF} 
The crystal-field (CF) level scheme for the $^7F_6$ multiplet of Tb$^{3+}$ ions in TbInO$_3$. 
The ground state, not listed in the table, is a doublet (d1); 
the excited states include five singlets (s1-s5) and three doublets (d2-d4). 
The energy and half width at half maximum (HWHM) are determined from Raman spectra measured at 5\,K. 
Units are meV.}
\begin{ruledtabular}
\begin{tabular}{ccccccccc}
Symmetry & s1  & s2  & s3  & s4  & d2  & d3  & s5   & d4   \\
\hline 
Energy   & 0.8 & 2.5 & 3.7 & 5.7 & 6.3 & 8.7 & 10.8 & 11.5 \\
HWHM     & 0.25 & 0.4 & 0.4 & 0.65 & 0.65 & 0.7 & 0.6  & 1.3  \\
\end{tabular}
\end{ruledtabular}
\end{table}

By now the low-energy CF level scheme is established [Table\,\ref{table:CF}], and we can relate our results to the two excitation branches measured in the INS study~\cite{Kiryukhin2019}: their gapless excitation extending to 1.4\,meV likely corresponds to the Raman-measured 0.8\,meV singlet with 0.5\,meV width; their broad excitation between 1.6 and 3.0\,meV matches well the Raman-measured 2.5\,meV singlet with 0.7\,meV width.

Low-temperature CF excitations in insulators are expected to be sharp~\cite{Cardona2000}. 
However, the linewidths of the low-energy CF excitations in TbInO$_3$ are much broader than the typical. 
To identify the reason for this broadening, first we exclude the effect of thermal fluctuations: the observed widths measured by INS and Raman are consistent even though the INS data are taken at 0.2\,K while the Raman data are acquired at 5\,K; moreover, the Raman linewidth of the s4 and d2 CF modes remains essentially same between 5 and 15\,K. 
Second, the same excitation branches measured by INS in TbIn$_{0.95}$Mn$_{0.05}$O$_3$ have very similar linewidth~\cite{Kiryukhin2019}, indicating that the broadening is not due to structural imperfection; 
in addition, the fact that the CF modes exhibit a Lorentzian lineshape indicates lack of inhomogeneous broadening. 
We suggest that the anomalously large CF linewidth of TbInO$_3$ is caused by
magnetic fluctuations near the SL ground state: the dynamics of the correlated yet non-ordered magnetic moments manifests itself through the width of CF excitations.

Thus, in this section we have shown that both Tb1 and Tb2 ions have essentially the same CF level scheme within the $^7F_6$ ground multiplet; particularly, their ground state is a non-Kramers doublet. Such a doublet ground state allows a finite magnetic moment, and the same CF level scheme for all Tb$^{3+}$ ions further supports the scenario of triangular magnetic lattice. We have suggested that the broad CF linewidth is related to the magnetic fluctuations resulting from the SL dynamics.

\section{Phonon modes\label{sec:P}}

In this section we identify the Raman-active phonon modes from the low-temperature spectra. 
We start from group-theoretical analyses to count the number and symmetries of Raman modes. 
In the high-temperature \textit{paraelectric} phase (space group $P6_3/mmc$, No. 194; point group D$_{6h}$), TbInO$_3$ has 10 atoms in one unit cell, and 5 Raman-active optical phonon modes: 1$A_{1g}\oplus 1E_{1g}\oplus 3E_{2g}$;
in the low-temperature \textit{ferroelectric} phase (space group $P6_3cm$, No. 185; point group C$_{6v}$), TbInO$_3$ has 30 atoms in one unit cell, and 38 Raman-active optical phonon modes: 9$A_{1}\oplus 14E_{1}\oplus 15E_{2}$ [Appendix~\ref{sec:AP}].

\begin{figure*}
\includegraphics[width=0.9\textwidth]{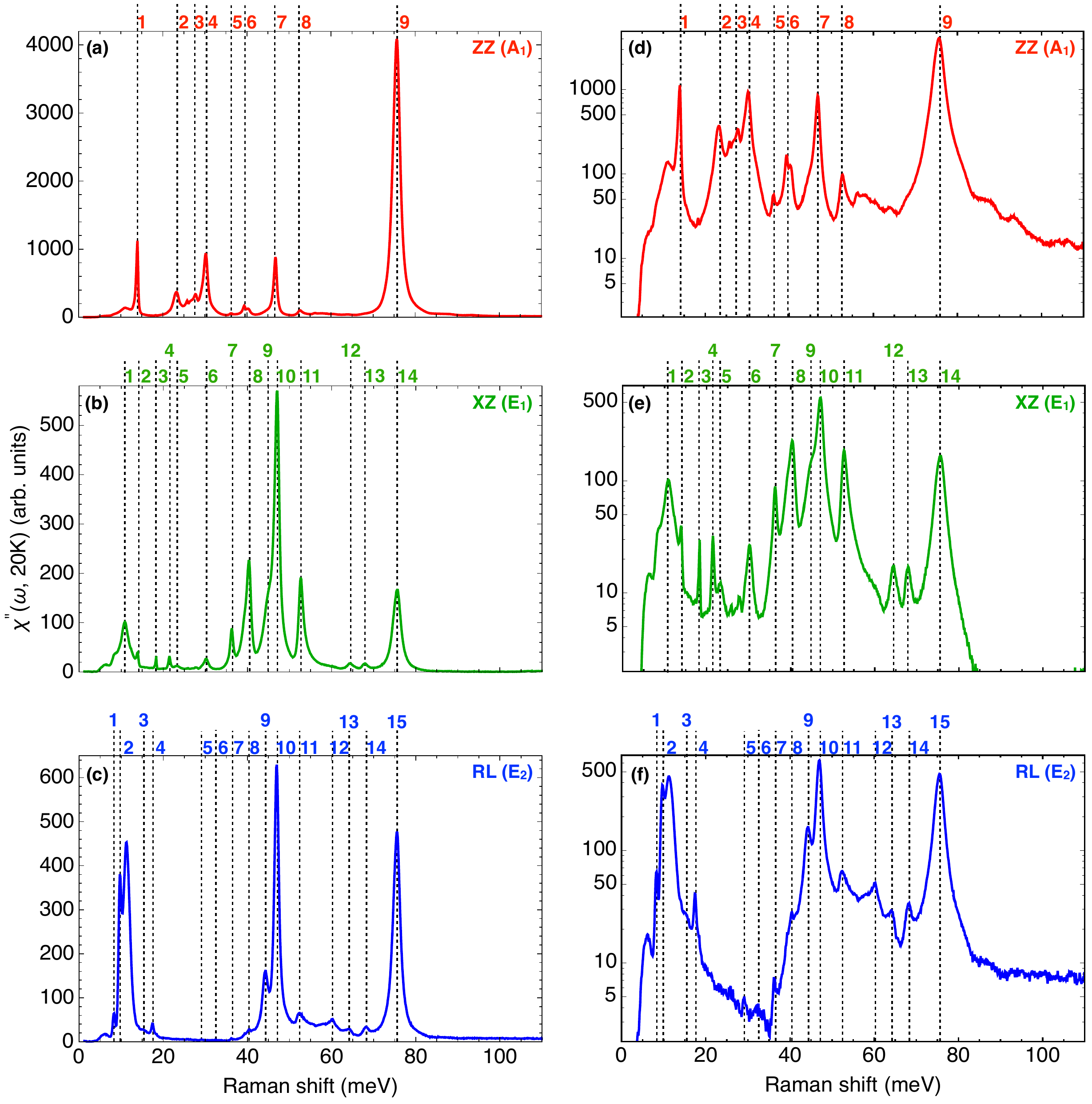}
\caption{\label{fig:P}Raman response $\chi''(\omega,20\,K)$ of the (a) A$_{1}$-symmetry, (b) E$_{1}$-symmetry and (c) E$_{2}$-symmetry phonon modes measured at 20\,K with 647\,nm excitation, plotted in linear scale. (d-f) The respective plots in semi-log scale. Unlabelled spectral features below 15\,meV are related to crystal-field modes; weak features at higher energy result from second-order scattering.}
\end{figure*}

The spectra of phonon modes at low temperature are presented in Fig.~\ref{fig:P} with both linear and semi-log scale. Unlabelled spectral features below 15\,meV are related to crystal-field modes; some weak features at higher energy, for example those in between A$_{1}^{(8)}$ and A$_{1}^{(9)}$ modes, result from second-order phonon scattering. 
In such scattering process, two phonons of zero total momentum are excited simultaneously; 
because the resulting spectral feature depends on the phonon dispersion and density of states, it does not have Lorentzian lineshape and commonly exhibits broad linewidth~\cite{Klein1981,Mai2019}.

The energies of the phonon modes at 20\,K are summarized in Table~\ref{table:P}. The energy of 9 phonon modes was identified by unpolarized Raman measurements of a series of hexagonal rare-earth (RE) REInO$_{3}$ compounds~\cite{Barnita2016}; these energy values are consistent with the result of this study.

\begin{table}[b]
\caption{\label{table:P}The energies of the A$_{1}$-symmetry, E$_{1}$-symmetry and E$_{2}$-symmetry Raman-active optical phonon modes at 20\,K. Units are meV.}
\begin{ruledtabular}
\begin{tabular}{cccc}
Number&A$_{1}$ modes&E$_{1}$ modes&E$_{2}$ modes\\
\hline 
1&13.9&10.2&8.1\\
2&23.1&13.8&9.6\\
3&26.8&18.3&15.2\\
4&30.1&21.4&17.4\\
5&36.2&23.1&29.0\\
6&39.7&30.2&32.0\\
7&46.7&36.2&36.1\\
8&52.6&40.2&40.3\\
9&75.6&44.9&44.0\\
10& &47.0&46.7\\
11& &52.5&52.2\\
12& &64.3&60.1\\
13& &67.8&64.1\\
14& &75.5&68.0\\
15& & &75.4\\
\end{tabular}
\end{ruledtabular}
\end{table}

The atomic displacements of a Raman-active optical phonon mode modulate the macroscopic polarizability. The stronger the induced polarizability, the stronger the intensity of the corresponding Raman mode. 
For TbInO$_3$, the paraelectric to ferroelectric transition not only causes some Raman-inactive modes in the paraelectric phase to become Raman-active in the ferroelectric phase, but also results in new Raman-active modes in the ferroelectric phase which have no correspondence to the modes in the paraelectric phase. 
Because the ferroelectric structure differs by only small distortions from the paraelectric structure, these additional Raman-active modes are expected to render only weak modulation of polarizability and in turn have weak intensity. 
On the contrary, the A$_{1}^{(9)}$, E$_{1}^{(10)}$, E$_{2}^{(2)}$, E$_{2}^{(10)}$ and E$_{2}^{(15)}$ modes have much stronger intensity than the other modes. These modes correspond to the 5 Raman-active modes of the paraelectric phase:
\begin{equation}
\begin{split}
A_{1}^{(9)} & \rightarrow A_{1g}; \\
E_{1}^{(10)} & \rightarrow E_{1g}; \\
E_{2}^{(2)}, E_{2}^{(10)}, E_{2}^{(15)} & \rightarrow 3E_{2g}.
\end{split}
\label{eq:modes}
\end{equation}

Hitherto, we have characterized the symmetry and measured the energy of all 38 Raman-active phonon modes at 20\,K. Moreover, we have interpreted the relative intensity of these modes in view of the structural change between the paraelectric and ferroelectric phases: the modes which are Raman-active in the paraelectric phase have strong intensity in the ferroelectric phase, while other modes have weak intensity.

\section{Vibronic modes in E$_{2}$ symmetry channel \label{sec:Vib}}

\begin{figure}[t]
\includegraphics[width=0.46\textwidth]{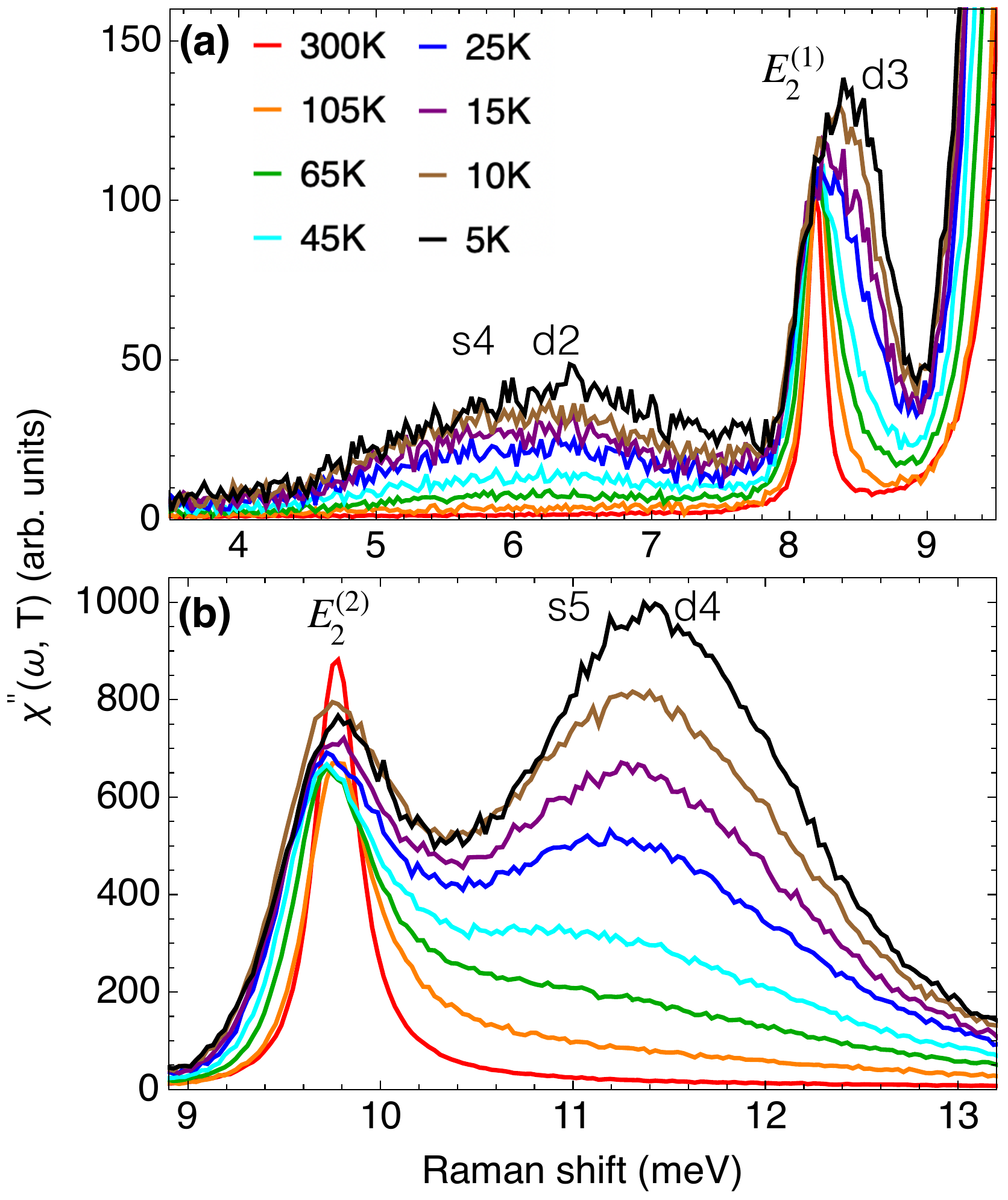}
\caption{\label{fig:Temp}Temperature dependence of the Raman response $\chi''(\omega,T)$ for (a) E$_{2}^{(1)}$ and (b) E$_{2}^{(2)}$ phonon modes, measured in RL scattering geometry with 676\,nm excitation. The CF modes are labelled by their respective final state; their bare responses are shown in Fig.\,\ref{fig:CF2}.}
\end{figure}

After presenting the CF and phonon spectra respectively in Sec.~\ref{sec:CF} and Sec.~\ref{sec:P}, we examine the hybrid vibronic spectral features resulting from the coupling between the CF and phonon modes. 
In Fig.\,\ref{fig:Temp} we present the temperature dependence of the E$_{2}^{(1)}$ and E$_{2}^{(2)}$ phonon modes. The apparent linewidth of the two modes anomalously increase on cooling. 
Moreover, the energy and HWHM of the CF modes, determined from XZ and ZZ spectra, cannot be directly used to fit the modes between 8 and 12\,meV in the RL spectrum. These two phenomena indicate presence of coupling between the two phonon modes and three CF modes in the 8-12\,meV energy range. 
Even above 100\,K, when the CF modes become very broad, the coupling effects remain noticeable, as it seen from the asymmetric lineshape of the E$_{2}^{(2)}$ mode [Fig.~\ref{fig:Temp}(b)].

Because of the close proximity of CF transitions to phonon modes, vibronic excitations have been observed in several rare-earth compounds, e.g. in Ce$_2$O$_3$~\cite{Cooper2019}, NdBa$_2$Cu$_3$O$_7$~\cite{Cardona1991}, and Ho$_2$Ti$_2$O$_7$~\cite{Gaulin2018}. 
In these cases, one CF transition interacts with one phonon mode, resulting in two vibronic states~\cite{Thalmeier1982}.

The E$_{2}$-symmetry spectra of TbInO$_3$ are more involved, because three CF transitions and two phonon modes are close in frequency, leading to complex spectra structure. 
Before analyzing this multi-mode problem, we consider a simplified case in which one phonon mode interacts with one CF mode. We model the bare response of the CF and phonon modes, in the absence of interaction, by Lorentzian function~\footnote{We note that a proper Raman response should satisfy the requirement $\chi^{\prime\prime}(\omega=0)=0$; only for $\omega \ll 0$ Eq.\eqref{eq:Lorentz} can be used as an approximation.}:
\begin{equation}
\chi^{\prime\prime}_{e,p}(\omega)=\frac{t_{e,p}^2\gamma_{e,p}}{(\omega-\omega_{e,p})^2+\gamma_{e,p}^2}\,, 
\label{eq:Lorentz}
\end{equation}
where the subscript "e" and "p" label the CF and phononic responses respectively; $t_{e,p}$ is the light-scattering vertex, $\omega_{e,p}$ is the mode's bare frequency, and $\gamma_{e,p}$ is the bare half width at half maximum (HWHM). % is related to the oscillator lifetime.

%\mpar{I think, the case you present in Fig.\,\ref{fig:CFS} is not the most relevant/clarifying. In particular, the anti-resonance feature is not at all obvious. Also, it is quite crowded. Why don't you consider case relevant to the E2(1)-d3 pair with varying the interaction strength. }
The interaction between the phonon and CF modes couples these dynamical responses; as a result, not only are the bare responses [Eq.\eqref{eq:Lorentz}] renormalized, but also an interference term appears in the total Raman response. The full Raman response in turn can be broken into three contributions:
\begin{equation}
\chi^{\prime\prime}(\omega)=\chi^{\prime\prime}_{pv}(\omega)+\chi^{\prime\prime}_{ev}(\omega)+\chi^{\prime\prime}_{int}(\omega)\,,
\label{eq:Contributions}
\end{equation}
in which the subscript "v" indicates presence of the interaction. The first two terms correspond to the phonon response proportional to square of light coupling vertex $t_p^2$ and the CF response proportional to $t_e^2$, respectively, while the third one, that is proportional to the $t_p t_e$ combination, is the interference term appearing due to the exciton-phonon coupling with strength $v$.

The exact expressions for the three contributions are given in the Appendix\,\ref{sec:AVibSimple}.
To illustrate the essential features, we consider the weak-coupling limit $v/\Delta\omega \ll 1$ with $\Delta\omega = \omega_p - \omega_e$ being the frequency difference between the phonon and CF modes. In this limit, the renormalized phonon response can be written in a Lorentzian form:
\begin{equation}
\chi^{\prime\prime}_{pv}(\omega)=\frac{t_p^2\gamma_{pv}}{(\omega-\omega_{pv})^2+\gamma_{pv}^2}\,.
\label{eq:P}
\end{equation}
The renormalization shifts the phonon frequency from $\omega_p$ to $\omega_{pv} = \omega_p + v^2\Delta\omega/(\gamma_e^{2}+\Delta\omega^2)$, and broadens the phonon HWHM from $\gamma_p$ to $\gamma_{pv} = \gamma_p + v^2\gamma_e/(\gamma_e^{2}+\Delta\omega^2)$. The simultaneous change of both frequency and HWHM is consistent with the general effect of interaction on the self energy of a state: the interaction influences both the real part of the self energy, which shifts the frequency, and the imaginary part of the self energy, which broadens the linewidth.

Because the bare responses of the phonon and CF modes have the same form [Eq.\eqref{eq:Lorentz}], the renormalized CF response also has a Lorentzian form, as required by symmetry. The energy is shifted to $\omega_{ev} = \omega_e - v^2\Delta\omega/(\gamma_p^{2}+\Delta\omega^2)$, and the HWHM is broadened to $\gamma_{ev} = \gamma_e + v^2\gamma_p/(\gamma_p^{2}+\Delta\omega^2)$. If $\omega_e > \omega_p$, $\Delta\omega$ is negative; the maximum of CF response is moved to higher energy and that of phononic response is moved to lower energy, characteristic of the conventional level-repulsion behavior.

The interference term has the following expression in the weak-coupling limit:
\begin{equation}
\chi^{\prime\prime}_{int}(\omega)=\frac{-2t_pt_ev[\gamma_p(\omega-\omega_e)+\gamma_e(\omega-\omega_p)]}{[(\omega-\omega_e)^2+\gamma_e^2][(\omega-\omega_p)^2+\gamma_p^2]}\,.
\label{eq:Int}
\end{equation}
The sign of this term depends not only on $v$, but also on phase difference between the CF and phonon oscillators. 
Because the phase of driven by light oscillator is flipping to the opposite one at the resonant frequency, the sign of this term changes between the bare CF frequency and the bare phonon frequency, close to the one which has smaller bare linewidth.
%Depending on the sign of $v$, the peak is skewed to the left or to the right of the original phonon frequency.

After discussing the basic properties of the coupling effect, we fit the E$_{2}$-symmetry spectra of TbInO$_3$ with the model described in Appendix~\ref{sec:AVib}. We use the energy and HWHM in Table~\ref{table:CF} for the three CF transitions; the energy of the two phonon modes is constrained to be no less than their values at 300\,K, and the HWHM is constrained to be no more than their values at 300\,K. From the fitting we obtain the values of the coupling constants listed in Table~\ref{table:C}.

\begin{table}[b]
\caption{\label{table:C}
The coupling strength ($v$), frequency difference ($\Delta\omega$), and dimensionless coupling constant ($v/\Delta\omega$) for interactions between E$_{2}^{(i)}$ phonons ($i$=1,2) and the CF transitions to the states d3, s5 and d4. Values are obtained by fitting the vibronic excitations.}
\begin{ruledtabular}
\begin{tabular}{ccccccc}
Pair & E$_{2}^{(1)}$-d3 & E$_{2}^{(1)}$-s5 & E$_{2}^{(1)}$-d4 & E$_{2}^{(2)}$-d3 & E$_{2}^{(2)}$-s5 & E$_{2}^{(2)}$-d4 \\ 
\hline 
$v$ (meV) & -0.31 & -0.35 & -0.46 & -0.60 & -0.68 & -0.91 \\
$\Delta\omega$ (meV) & -0.44 & -2.59 & -3.24 & 1.38 & -0.77 & -1.42 \\
$v/\Delta\omega$ & 0.69 & 0.13 & 0.14 & -0.44 & 0.89 & 0.64 \\
\end{tabular}
\end{ruledtabular}
\end{table}

\begin{figure}[t]
\includegraphics[width=0.48\textwidth]{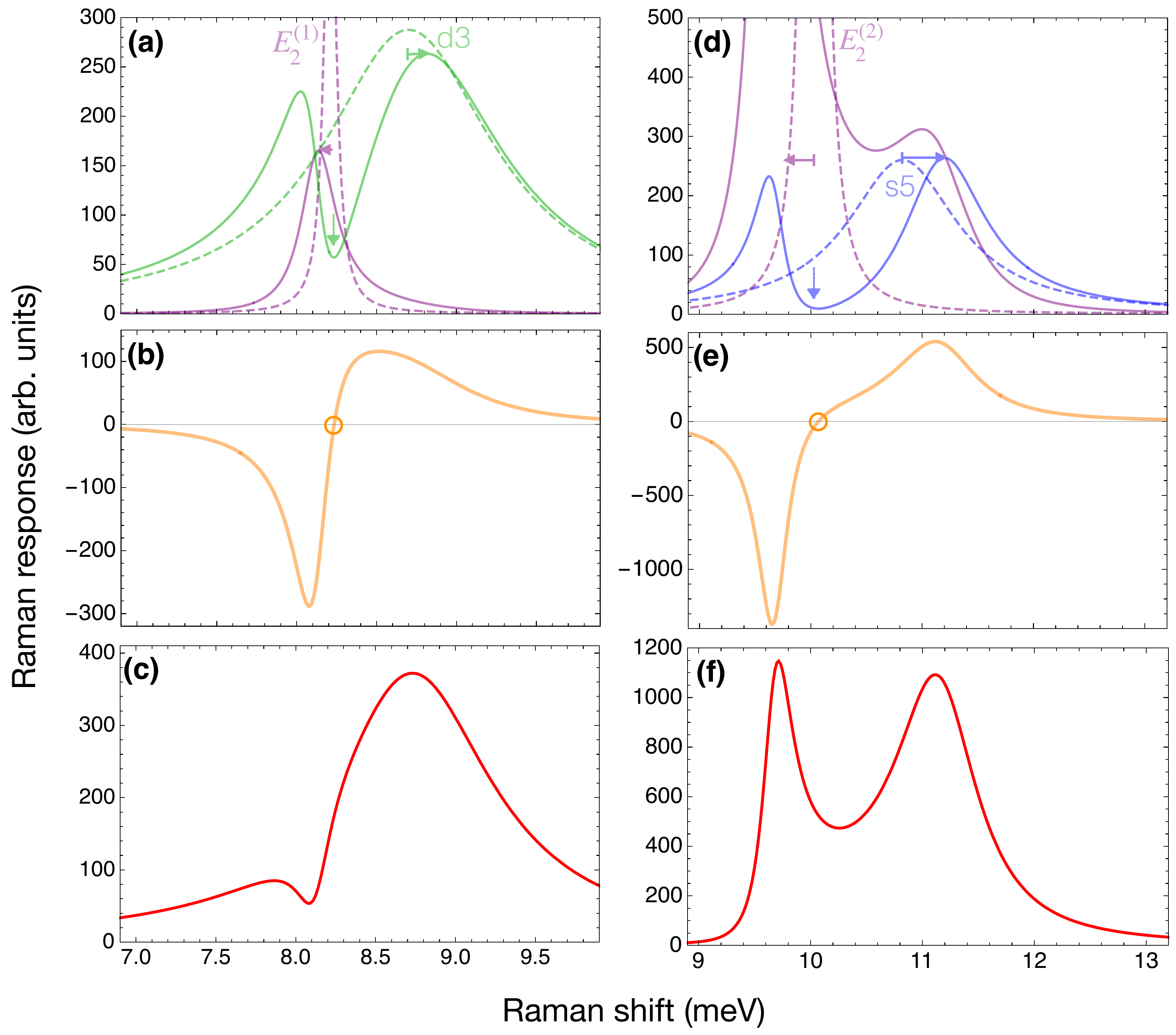}
\caption{\label{fig:CFI} 
Illustration for coupling between (a-c) the E$_{2}^{(1)}$ phonon and d3 CF modes; (d-f) the E$_{2}^{(2)}$ phonon and s5 CF modes.
(a) and (d): The Raman response functions for bare $\chi^{\prime\prime}_{e,p}(\omega)$ (dashed lines) and renormalized by interaction $\chi^{\prime\prime}_{ev,pv}(\omega)$ (solid lines). The shift of frequency is labeled by horizontal arrows; the anti-resonant feature of $\chi^{\prime\prime}_{ev}(\omega)$ at the bare phonon frequency is labeled by vertical arrows.
(b) and (e): The interference term $\chi^{\prime\prime}_{int}(\omega)$. The bare phonon frequency, at which $\chi^{\prime\prime}_{int}(\omega)$ changes sign, is labelled by a small circle.
(c) and (f): The total Raman response $\chi^{\prime\prime}(\omega)$ [Eq.\eqref{eq:Contributions}], corresponds to the sum of the renormalized responses and the interference term.}
\end{figure}
From Table~\ref{table:C}, we notice that the E$_{2}^{(1)}$ phonon mainly couples to the d3 CF transition, and the E$_{2}^{(2)}$ phonon has the strongest dimensionless coupling to the s5 CF transition. 
%\mpar{The Fig.\,\ref{fig:CFI} is still quite messy to digest.}
We study these two pairs separately in Fig.\,\ref{fig:CFI}. First we discuss the renormalization effect. Because $v/\Delta\omega \approx 1$, the weak-coupling limit does not apply to these two pairs and the renormalized responses do not have a Lorentzian lineshape [Fig.\,\ref{fig:CFI}(a) and (d)]. However, the level-repulsion behavior and the broadening of lineshape seen in the weak-coupling case still applies in the strong-coupling case. 
At the bare phonon frequency, the renormalized CF response exhibit an anti-resonant spectral feature, which arises due to the destructive interference of the two modes. The new local maximum near the anti-resonance results instead from constructive interference. 
Second, we consider the interference term, which changes sign essentially at the bare phonon frequency [Fig.\,\ref{fig:CFI}(b) and (e)] because the phonon mode is much sharper than the CF mode, consistent with discussion for the weak-coupling case. 
Third, we examine the total response, the sum of the renormalized responses and the interference term [Fig.\,\ref{fig:CFI}(c) and (f)]. 
Because of the much smaller relative intensity of the E$_{2}^{(1)}$ mode, it is "absorbed" into a broad composite vibronic feature [Fig.\,\ref{fig:CFI}(c)]. The E$_{2}^{(2)}$ phonon, on the contrary, has a much larger relative to CF transitions intensity; therefore it remains identifiable in the vibronic spectra [Fig.\,\ref{fig:CFI}(f)].

\begin{figure}
\includegraphics[width=0.48\textwidth]{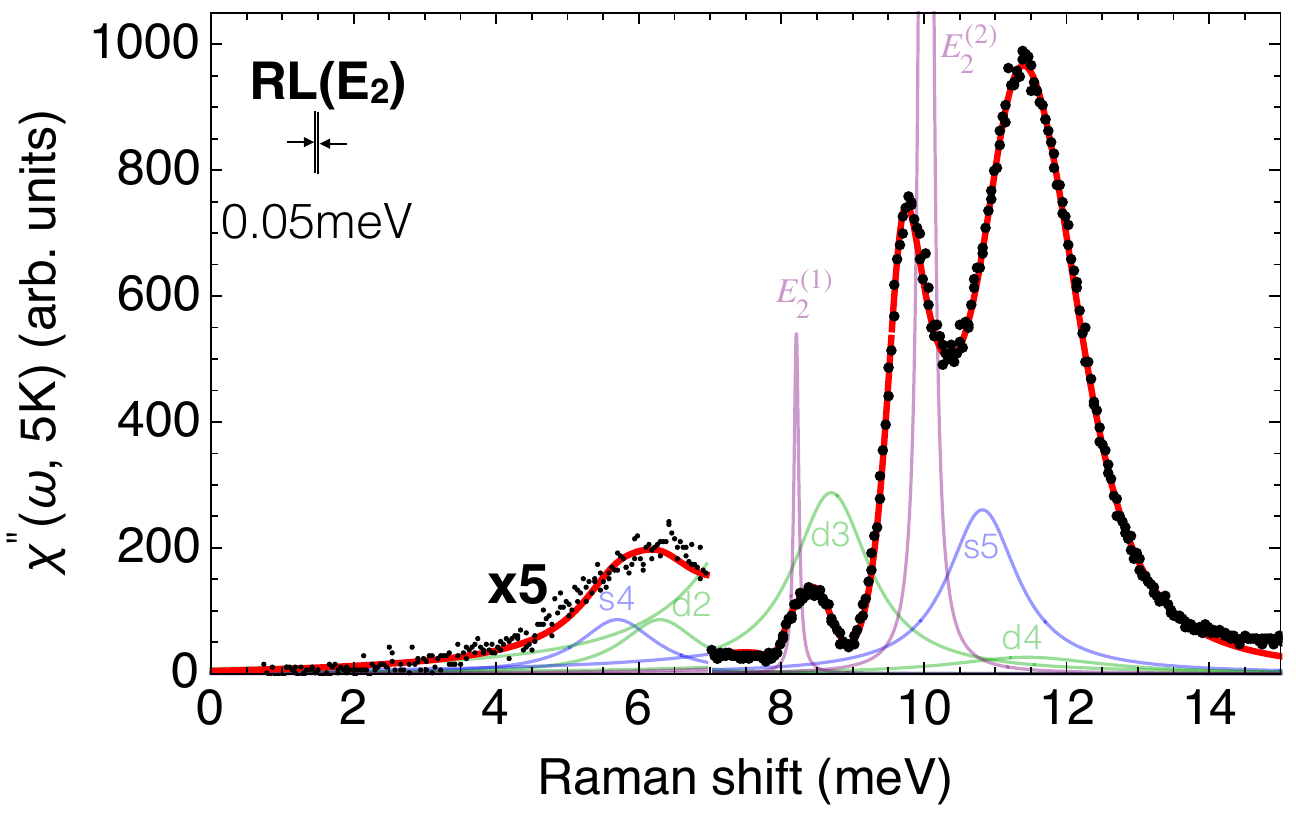}
\caption{\label{fig:CF2}The low-energy crystal-field (CF) transitions in TbInO$_3$, measured at 5\,K in RL scattering geometry.
The Raman data are represented by black dots. 
The red line represents the fit.
The blue lines show doublet-to-singlet CF oscillators labeled by the final singlet state, and the green lines show doublet-to-doublet CF transitions; 
the phonon modes, labeled by their symmetry, are shown in purple. 
The Raman data and fitting curves below 7\,meV are multiplied by a factor of 5 for clarity.}
\end{figure}

In Fig.\,\ref{fig:CF2} we present the overall effect of the coupling between three CF modes and two phonon modes. Although the E$_{2}^{(1)}$ phonon mainly couples to the d3 CF transition, the whole effect of coupling to three CF modes [Fig.\,\ref{fig:CF2}(a)] is similar to Fig.\,\ref{fig:CFI}(c), except for the disappearance of the weak dip near 8.1\,meV. However, for the E$_{2}^{(2)}$ phonon, the coupling to the d3 and d4 CF modes is not negligible; hence the whole effect is different from what is shown in Fig.\,\ref{fig:CFI}(f), especially the intensity.

Summing-up, in this section we identified the hybrid vibronic features from the temperature dependence of the E$_{2}^{(1)}$ and E$_{2}^{(2)}$ phonon modes. 
These two phonon modes couple to three CF modes in the same energy range. 
Such coupling results from the modulation of the electron-cloud distribution of the CF states by lattice vibration, and is facilitated by the energetic proximity of the CF and phonon modes. 
We illustrated the physics of this type of interaction by considering a one-by-one problem in the weak coupling limit. 
By fitting the spectrum of the vibronic modes, we have found that the coupling constant is comparable to the frequency difference between pairs of CF and phonon modes, indicating the strong-coupling regime. 
Presence of these vibronic excitations further implies strong spin-lattice dynamics. 

\section{Specific heat and entropy\label{sec:SH}}

In this section we demonstrate that the specific heat and entropy data are consistent with the CF level scheme determined in Sec.~\ref{sec:CF}. 
We use four components to model the specific heat of TbInO$_3$: the nuclear, CF, acoustic-phonon, and optical-phonon contributions [Appendix~\ref{sec:ASH}]. In particular, based on the Raman results we use Lorentzian lineshapes with the energy and width given in Table~\ref{table:CF} for the CF states. The CF contribution to the specific heat can be expressed in the following form:
\begin{multline}
C_E=\frac{R}{(k_BT)^2}\{\frac{1}{Z_E}\sum_{i}\int L_{i}(\epsilon)\epsilon^2e^{(-\epsilon/k_BT)}d\epsilon-\\
[\frac{1}{Z_E}\sum_{i}\int L_{i}(\epsilon)\epsilon e^{(-\epsilon/k_BT)}d\epsilon]^2\}~,
\label{eq:SHE}
\end{multline}
which is a generalization of the formula for CF levels with Dirac $\delta$ function as lineshape~\cite{Tari2003}. In Eq.~(\ref{eq:SHE}), $Z_E=\sum_{i}\int L_{i}(\epsilon)exp(-\epsilon/k_BT)d\epsilon$ is the partition function for the CF levels; $i$ (from 1 to 8) labels the individual CF energy levels, and L$_i$($\epsilon$) is the normalized Lorentzian function for energy level $i$.

\begin{figure}
\includegraphics[width=0.46\textwidth]{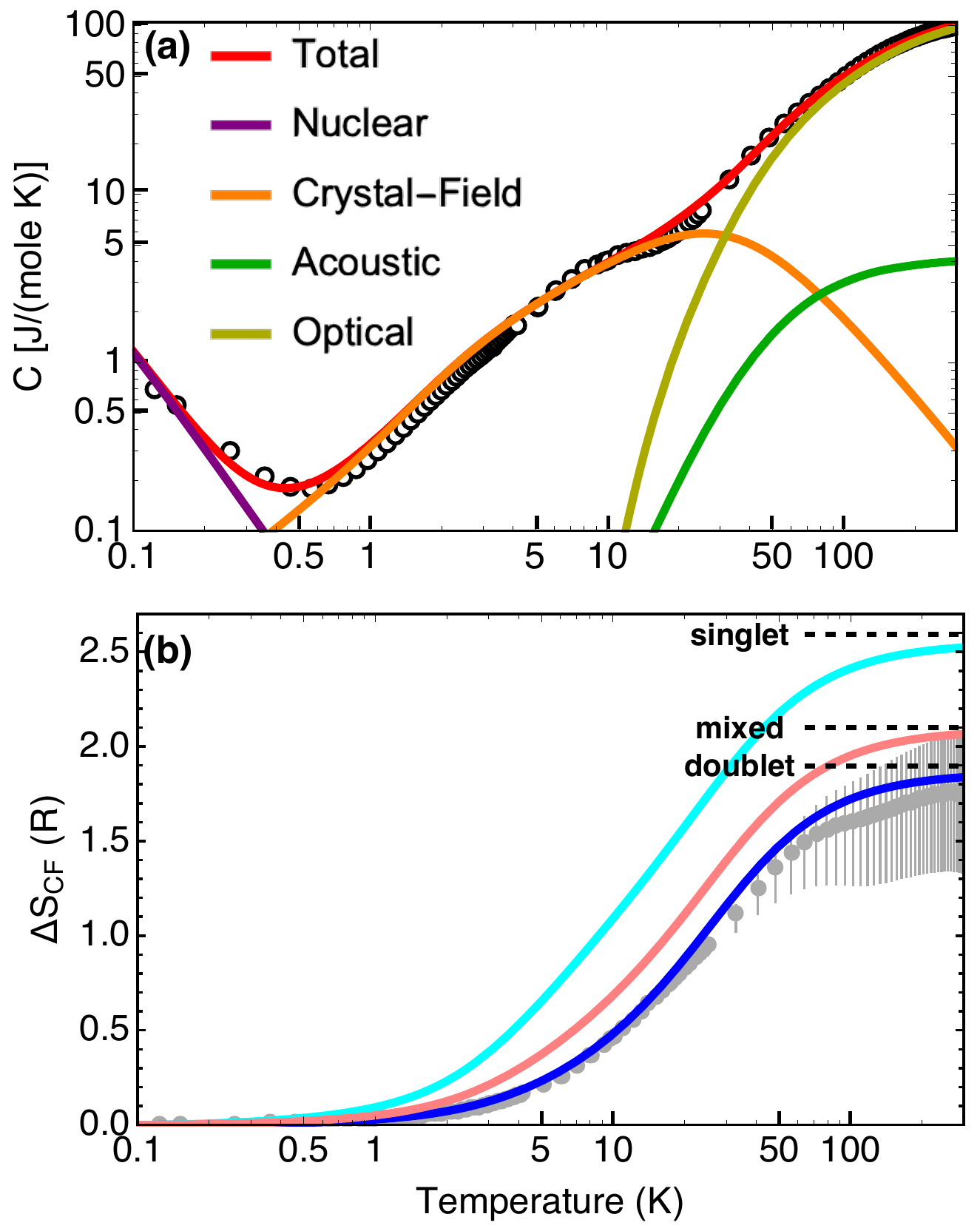}
\caption{\label{fig:SH}The specific heat and electronic entropy of TbInO$_3$. (a) Log-log plot of the specific heat data from Ref.~\cite{Cheong2019} (open black circles), and the calculated curves. The total specific heat composes of the nuclear, crystal-field, acoustic-phonon and optical-phonon contributions. (b) Semi-log plot of the entropy corresponding to the CF excitations (solid gray circles), and the calculated curves. The unit of entropy is the ideal gas constant $R=8.314$\,J/(mol K). The blue curve corresponds to the crystal-field level scheme in Table~\ref{table:CF}; the cyan curve assumes that the ground state is a singlet and the first excited state is a doublet; the pink curve in middle assumes that Tb1 has a singlet ground state and Tb2 has a doublet one.}
\end{figure}

In Fig.\,\ref{fig:SH}(a) we compare calculated and measured specific heat data~\cite{Cheong2019}. The specific heat below 0.5\,K can be accounted for by nuclear Schottky contribution, and the high-temperature specific heat is mostly contributed by phonon modes. In particular, the CF contribution agrees well with the data between 0.5\,K and 10\,K.

In the previous study~\cite{Cheong2019}, the low-temperature specific heat was fitted by assuming a singlet ground state for Tb1 ions and a doublet one for Tb2 ions. 
As a result, a large residual specific heat between 1 and 4\,K was shown after subtracting the nuclear and CF contributions. 
In~\cite{Cheong2019} authors model the CF states by Dirac $\delta$-function in energy, while we consider the finite width of these excitations. 
We propose that the contribution from magnetic fluctuations, which correspond to SL dynamics near the ground state, is effectively treated in the width of the CF states. 
Hence no residual specific heat should appear in our analysis. 

Regarding the entropy corresponding to the CF excitations, if the ground state is a doublet, the entropy should saturate at $R [\ln(2J+1)-\ln2]=1.9 R$ ($R$ is the ideal gas constant); and if it is a singlet, the entropy should saturate at $R \ln(2J+1) =2.6 R$. The entropy change $\Delta S$ from T$_1$ to T$_2$ can be calculated from the specific heat $C$ by integration:
\begin{equation}
\Delta S=\int_{T_1}^{T_2}\frac{C}{T}dT~.
\label{eq:Entropy}
\end{equation}
We first subtract the nuclear and phononic contributions from the experimental specific-heat data, and then use Eq.~(\ref{eq:Entropy}) to find the entropy corresponding to the CF excitations. The fact that the entropy at 300\,K is very close to $1.9 R$ further supports the assignment to doublet ground state [Fig.~\ref{fig:SH}(b)]. 
Moreover, an assumption that Tb1 and Tb2 both have a doublet ground state describes the entropy much better than an assumption that Tb1 has a singlet ground state and Tb2 has a doublet one.

In this section, we described the specific heat by the nuclear, CF, and phononic contributions. 
In particular, we have accounted the finite linewidth of the CF states as determined by spectroscopic data. 
We demonstarted that the temperature dependence of the CF entropy is consistent with a doublet ground state for both Tb$^{3+}$ ion sites. 

\section{Conclusions\label{sec:Con}}

In this work, we study the electronic and phononic excitations of TbInO$_3$ by polarization resolved Raman spectroscopy. 
We establish that TbInO$_3$ is hosting a triangular magnetic lattice of the non-Kramers doublet ground states. 
Specifically, we discuss inter-multiplet excitations within the $^7F$ term, intra-multiplet excitations within the $^7F_6$ ground multiplet, Raman-active phonon modes, specific heat data, and discover hybrid vibronic modes.

We observe the ${^5D}\,\rightarrow\,{^7F}$ luminescence continuum centered around 2\,eV, and the $^7F$ inter-multiplet Raman excitations. The energy of the $^7F$ multiplets is between 0.3 and 0.7\,eV. These multiplets exhibit at low temperature fine intra-multiplet CF structures which are clustered within their respective particular energy range.

We measure the CF excitations and in turn establish the CF level scheme for the $^7F_6$ ground multiplet. 
We show that both Tb1 and Tb2 ions, though having slightly different crystal environment, have essentially the same CF level scheme. 
In particular, they both exhibit a non-Kramers doublet ground state. 
This result supports the scenario of triangular magnetic lattice, which could support U(1)-symmetry SL ground state. 
Moreover, the orbital-degenerate ground state could result in spin-orbit-entangled multipolar phases. 
These possibilities renders TbInO$_3$ as a suitable platform to explore the SL physics. 
The enhanced linewidth of the CF states further serves as an indirect evidence for the magnetic fluctuations originating from SL dynamics.

We determine the energies of all 38 Raman-active phonon modes: 9$A_{1}\oplus 14E_{1}\oplus 15E_{2}$ and identify the five modes which remain Raman-active in the high-temperature paraelectric phase

We discover hybrid vibronic excitations composed of coupled CF and phonon modes in the quadrupolar E$_{2}$ symmetry channel. 
Such coupling results from the modulation of the electron-cloud distribution of the CF states by lattice vibration, and is facilitated by the energetic overlap between the CF and phonon modes. 
We use a simple one-by-one model to illustrate the physics behind this type of interaction, including level repulsion, linewidth broadening, and spectral interference effect. 
We use a model fit to the E$_{2}$-symmetry spectrum and derive the electron-phonon coupling strength. 
Because the energetic proximity of CF and phonon modes is common in rare-earth compounds, the developed method for decomposing Raman spectra of strongly-coupled phononic and CF excitations may have wide applicability. 

We also fit the specific-heat data by the sum of nuclear, CF and phononic contributions consistent with obtained spectroscopic parameters. 
The temperature dependence of the CF entropy further supports the CF level scheme determined from Raman measurements, therefore, may be of interest for further exploration of SL physics.

\begin{acknowledgments}
We thank V.\,Kriyukhin, M.G.\,Kim, and Y.-B.\,Li for discussions. 
The spectroscopic work at Rutgers (M.Y., X.W., and G.B.) was supported by the NSF under Grant DMR-1709161. 
The TbInO$_3$ crystal growth and characterization (X.X., J.K., and S.-W.C.) was supported by the DOE under Grant No.\,DE-FG02-07ER46382. 
The work at NICPB was supported by the Estonian Research Council Grant No. PRG736 and 
the European Research Council (ERC) under Grant Agreement No.\,885413.
\end{acknowledgments}

\appendix

\section{Multiplet splitting induced by crystal-field potential\label{sec:AM}}

The multiplets of the $^7F$ term are split by the CF potential. In Table~\ref{table:M} we classify the symmetry of the resulting fine levels by the irreducible representations of C$_{3v}$ group. On reducing the site symmetry from C$_{3v}$ to C$_{3}$, the A$_{1}$ and A$_{2}$ states of C$_{3v}$ group mix into the A states of C$_{3}$. 
\begin{table}[b]
\caption{\label{table:M}
Splitting of the  $^7F_J$ multiplet by the C$_{3v}$-symmetry CF potential. The symmetry of the CF states is classified by the irreducible representations of C$_{3v}$ group.}
\begin{ruledtabular}
\begin{tabular}{lccccccc}
J value & 6 & 5 & 4 & 3 & 2 & 1 & 0 \\ \hline 
A$_{1}$ & 3 & 1 & 2 & 1 & 1 & 0 & 1 \\
A$_{2}$ & 2 & 2 & 1 & 2 & 0 & 1 & 0 \\
E       & 4 & 4 & 3 & 2 & 2 & 1 & 0 \\
\end{tabular}
\end{ruledtabular}
\end{table}

\section{Classification of the $\Gamma$-point phonons \label{sec:AP}}

\begin{figure}[b]
\includegraphics[width=0.48\textwidth]{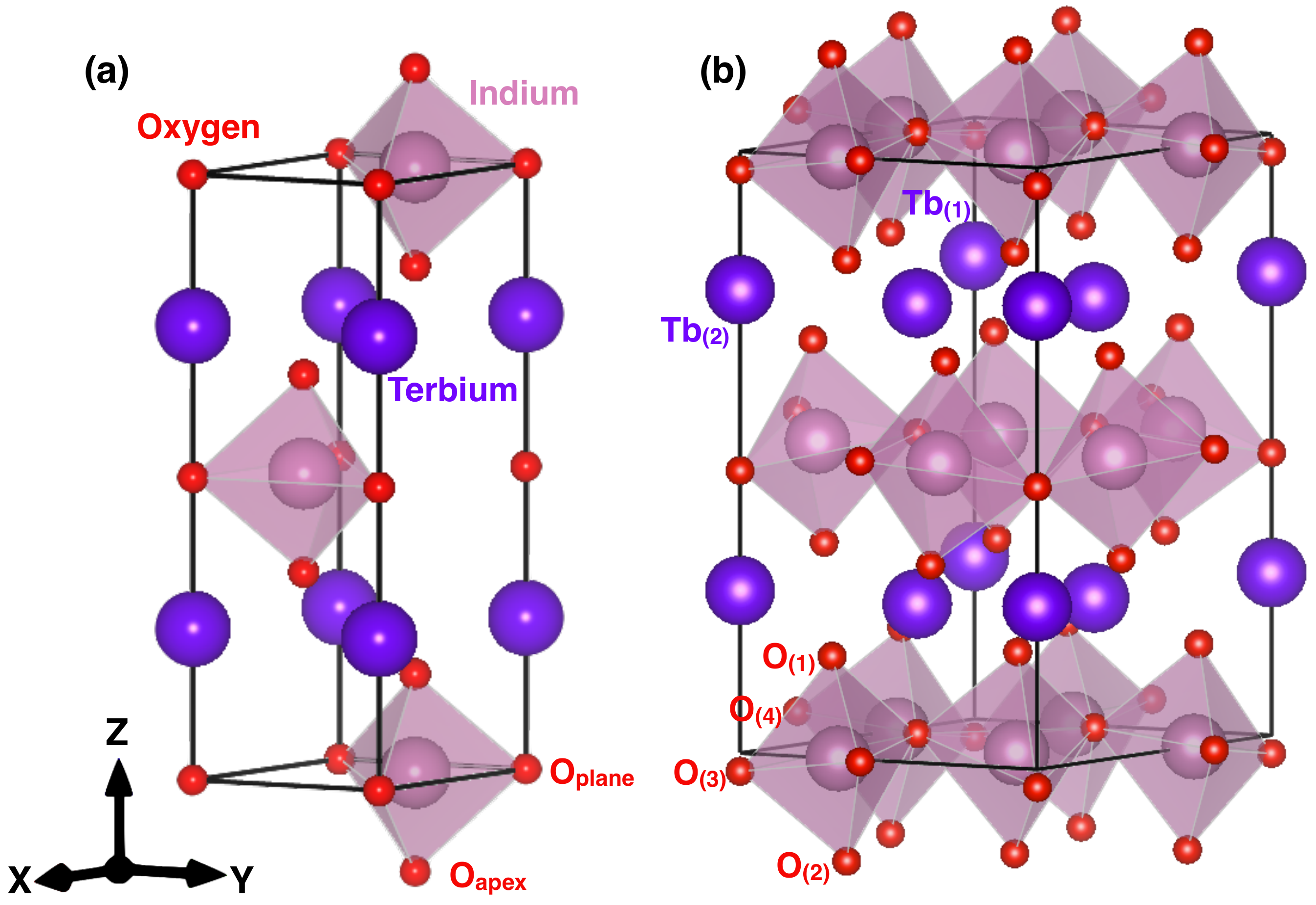}
\caption{\label{fig:TP}
Crystal structure for (a) paraelectric and (b) ferroelectric phases of TbInO$_3$.
The oxygen, indium and terbium atoms are shown in red, pink, and purple colors, respectively.
The black frame indicates the unit cell.
The oxygen sites of different symmetry are labelled in each case. Two Tb sites of different symmetry are labelled in panel (b).
}
\end{figure}

In Fig.~\ref{fig:TP} we compare the paraelectric and ferroelectric phases of TbInO$_3$. We first discuss the high-temperature paraelectric phase [Fig.~\ref{fig:S}(a)], with space group $P6_3/mmc$ and point group D$_{6h}$. Although high-temperature Raman measurements were not performed in this study, considering the paraelectric phase is helpful for understanding many aspects of the Raman-active phonons, particularly the large intensity difference for different modes. The unit cell contains two formula units, having two in-equivalent oxygen sites (O$_{\textrm{apex}}$ and O$_{\textrm{plane}}$) and only one site for Tb and In. The Tb site is the center of inversion and Tb atoms therefore cannot participate in Raman-active lattice vibrations. The $\Gamma$-point phonons are classified in Table~\ref{table:Para}; there is a total of 5 Raman-active phonon modes.

\begin{table*}
\caption{\label{table:Para}Classification of the $\Gamma$-point phonons for the high-temperature \textit{paraelectric} TbInO$_3$ (space group $P6_3/mmc$, No. 194; point group D$_{6h}$; Z=2).}
\begin{ruledtabular}
\begin{tabular}{lccc}
     & Wyckoff  & Site     & Irreducible     \\
Atom & notation & symmetry & representations \\
\hline
Tb   & 2(a)     & D$_{3d}$ & A$_{2u}$+B$_{2u}$+E$_{1u}$+E$_{2u}$ \\
In   & 2(c)     & D$_{3h}$ & A$_{2u}$+B$_{1g}$+E$_{1u}$+E$_{2g}$ \\
O$_{\textrm{plane}}$ & 2(b) & D$_{3h}$ & A$_{2u}$+B$_{1g}$+E$_{1u}$+E$_{2g}$ \\
O$_{\textrm{apex}}$  & 4(f) & C$_{3v}$ & A$_{1g}$+A$_{2u}$+B$_{1g}$+B$_{2u}$+E$_{1g}$+E$_{1u}$+E$_{2g}$+E$_{2u}$ \\
\hline
\multicolumn{4}{c}{$\Gamma$-point phonons}\\
\multicolumn{3}{c}{$\Gamma_{\textrm{Raman}}$=A$_{1g}$+E$_{1g}$+3E$_{2g}$}&\multicolumn{1}{c}{$\Gamma_{\textrm{acoustic}}$=A$_{2u}$+E$_{1u}$}\\
\multicolumn{3}{c}{$\Gamma_{\textrm{IR}}$=3A$_{2u}$+3E$_{1u}$}&\multicolumn{1}{c}{$\Gamma_{\textrm{silent}}$=3B$_{1g}$+2B$_{2u}$+2E$_{2u}$}\\
\end{tabular}
\end{ruledtabular}
\vspace{-3mm}
\end{table*}
\begin{table*}
\caption{\label{table:Ferro}Classification of the $\Gamma$-point phonons for the low-temperature \textit{ferroelectric} TbInO$_3$ (space group $P6_3cm$, No. 185; point group C$_{6v}$); Z=6.}
\begin{ruledtabular}
\begin{tabular}{lccc}
     & Wyckoff  & Site     & Irreducible     \\
Atom & notation & symmetry & representations \\
\hline
Tb$_{(1)}$   & 2(a)     & C$_{3v}$ & A$_{1}$+B$_{1}$+E$_{1}$+E$_{2}$ \\
Tb$_{(2)}$   & 4(b)     & C$_{3}$  & A$_{1}$+A$_{2}$+B$_{1}$+B$_{2}$+2E$_{1}$+2E$_{2}$ \\
In           & 6(c)     & C$_{s}$  & 2A$_{1}$+A$_{2}$+2B$_{1}$+B$_{2}$+3E$_{1}$+3E$_{2}$ \\
O$_{(1)}$    & 6(c)     & C$_{s}$  & 2A$_{1}$+A$_{2}$+2B$_{1}$+B$_{2}$+3E$_{1}$+3E$_{2}$ \\
O$_{(2)}$    & 6(c)     & C$_{s}$  & 2A$_{1}$+A$_{2}$+2B$_{1}$+B$_{2}$+3E$_{1}$+3E$_{2}$ \\
O$_{(3)}$    & 2(a)     & C$_{3v}$ & A$_{1}$+B$_{1}$+E$_{1}$+E$_{2}$ \\
O$_{(4)}$    & 4(b)     & C$_{3}$  & A$_{1}$+A$_{2}$+B$_{1}$+B$_{2}$+2E$_{1}$+2E$_{2}$ \\
\hline
\multicolumn{4}{c}{$\Gamma$-point phonons}\\
\multicolumn{3}{c}{$\Gamma_{\textrm{Raman}}$=9A$_{1}$+14E$_{1}$+15E$_{2}$}&\multicolumn{1}{c}{$\Gamma_{\textrm{acoustic}}$=A$_{1}$+E$_{1}$}\\
\multicolumn{3}{c}{$\Gamma_{\textrm{IR}}$=9A$_{1}$+14E$_{1}$}&\multicolumn{1}{c}{$\Gamma_{\textrm{silent}}$=5A$_{2}$+10B$_{1}$+5B$_{2}$}\\
\end{tabular}
\end{ruledtabular}
\vspace{-3mm}
\end{table*}

The low-temperature ferroelectric phase [Fig.~\ref{fig:S}(b)], with space group $P6_3cm$ and point group C$_{6v}$, can be obtained from the paraelectric phase by tilting the InO$_5$ bi-pyramids and buckling the Tb layers~\cite{Iliev1997}. Because of the tilting, the apex oxygen sites O$_{(1)}$ and O$_{(2)}$, and the planer oxygen sites O$_{(3)}$ and O$_{(4)}$ become non-equivalent; as a result of the buckling, Tb$_{(1)}$ site becomes slightly out of the plane formed by Tb$_{(2)}$. The unit cell contains six formula units. We note that the number of Raman-active modes should increase for two reasons: (i) the absence of central inversion symmetry renders the Raman- and infrared (IR)-active modes not mutually exclusive; (ii) the larger volume of the unit cell leads to new phonon modes at $\Gamma$ point, some of which are Raman-active. From Table~\ref{table:Ferro}, the number of Raman-active phonon modes increase from 5 to 38.

\section{Mathematical expressions for the coupling between one phonon and one crystal-field mode}\label{sec:AVibSimple}

In Sec.~\ref{sec:Vib} we consider a simplified case in which one phonon mode interacts to one CF mode; the weak-coupling limit is taken in order to illustrate the essential features. In this section, we present the exact expressions for the three contributions in Eq.\eqref{eq:Contributions}.

The renormalized phonon response:
\begin{equation}
\chi^{\prime\prime}_{pv}(\omega)=t_p^2\{\gamma_{p}[(\omega-\omega_{e})^2+\gamma_{e}^{2}]+\gamma_{e}v^2\}/D\,.
\label{eq:Sim1}
\end{equation}

The renormalized CF response:
\begin{equation}
\chi^{\prime\prime}_{ev}(\omega)=t_e^2\{\gamma_{e}[(\omega-\omega_{p})^2+\gamma_{p}^{2}]+\gamma_{p}v^2\}/D\,.
\label{eq:Sim2}
\end{equation}

The interference term:
\begin{equation}
\chi^{\prime\prime}_{int}(\omega)=-2t_pt_ev[\gamma_p(\omega-\omega_e)+\gamma_e(\omega-\omega_p)]/D\,.
\label{eq:Sim3}
\end{equation}

We note that $\chi^{\prime\prime}_{p}(\omega)$ and $\chi^{\prime\prime}_{e}(\omega)$ are related by the exchange of subscript $p \leftrightarrow e$ in their respective expressions, as required by symmetry. For the same reason, the interference term is invariant under the exchange of subscript $p \leftrightarrow e$. The three expressions have the same denominator $D$, which is
\begin{multline}
[\gamma_e^{2}+(\omega-\omega_e)^2][\gamma_p^{2}+(\omega-\omega_p)^2]\\
-2v^2[\gamma_e\gamma_p+(\omega-\omega_e)(\omega-\omega_p)]+v^4\,.
\label{eq:Sim4}
\end{multline}

\section{The fitting model for the vibronic features\label{sec:AVib}}

The apparent spectral linewidth of the E$_{2}^{(1)}$ and E$_{2}^{(2)}$ phonon modes anomalously increases on cooling [Fig.\,\ref{fig:Temp}], indicating existence of interaction between the phonon modes and the CF excitations. Such interaction originates from the modulation of the electron-cloud distribution of CF states by lattice vibration. We use a Green's function formalism to construct a model describing this physics~\cite{Mai2021}. In essence, the Raman response of the coupled system is calculated from the interacting Green's function:
\begin{equation}
\chi^{\prime\prime}\sim\Im T^{T}GT~,
\label{eq:coup1}
\end{equation}
in which $T$ denotes the vertices for light scattering process and $G$ is the Green's function for the interacting phononic and CF excitations. The Green's function $G$ can be obtained by solving the Dyson equation:
\begin{equation}
G=(G_0^{-1}-V)^{-1}~,
\label{eq:coup2}
\end{equation}
where $G_0$ is the bare Green's function and $V$ represents the interaction. To proceed, we first need to consider which phonon and CF modes are involved, and the way they couple to each other.

For C$_{3v}$/C$_{3}$ groups, the doublet-to-singlet transitions appear only in E symmetry channel; the doublet-to-doublet transitions appear both in A$_{1}$/A and in E symmetry channels. The E-symmetry component of the CF excitations can interact with the E$_{1}$- and E$_{2}$-symmetry phonon modes; the A$_{1}$/A-symmetry component of the CF excitations interacts with the A$_{1}$-symmetry phonon modes.

From the measured Raman spectra, we identify there CF modes in the energy range between 8 and 12\,meV: the first one is at 8.6\,meV (a peak in both XZ and RL spectra), the second one at 10.8\,meV (a peak in XZ spectrum), and the third one at 11.4\,meV (a peak in RL spectrum). Two of them are doublets and the rest one is a singlet. The three E-symmetry components of the CF excitations interact with two E$_{1}$-symmetry and two E$_{2}$-symmetry phonon modes; the two A$_{1}$/A-symmetry component of the CF excitations interact with one A$_{1}$-symmetry phonon mode.

For the E-symmetry components of the CF excitations, the bare Green's function $G_{E0}$ and the coupling interaction $V_{E}$ are
\begin{equation}
G_{E0}=
\begin{pmatrix}
G_{p1} & 0      & 0      & 0      & 0      & 0      & 0      \\
0      & G_{p2} & 0      & 0      & 0      & 0      & 0      \\
0      & 0      & G_{p3} & 0      & 0      & 0      & 0      \\
0      & 0      & 0      & G_{p4} & 0      & 0      & 0      \\
0      & 0      & 0      & 0      & G_{e1} & 0      & 0      \\
0      & 0      & 0      & 0      & 0      & G_{e2} & 0      \\
0      & 0      & 0      & 0      & 0      & 0      & G_{e3} \\ 
\end{pmatrix}~,
\label{eq:GE0}
\end{equation}
\begin{equation}
V_{E}=
\begin{pmatrix}
0      & 0      & 0      & 0      & v_{11} & v_{12} & v_{13} \\
0      & 0      & 0      & 0      & v_{21} & v_{22} & v_{23} \\
0      & 0      & 0      & 0      & v_{31} & v_{32} & v_{33} \\
0      & 0      & 0      & 0      & v_{41} & v_{42} & v_{43} \\
v_{11} & v_{21} & v_{31} & v_{41} & 0      & 0      & 0      \\
v_{12} & v_{22} & v_{32} & v_{42} & 0      & 0      & 0      \\
v_{13} & v_{23} & v_{33} & v_{43} & 0      & 0      & 0      \\
\end{pmatrix}~.
\label{eq:VE}
\end{equation}
The phononic Green's function $G_{pi}$ (i=1,2) corresponds to the E$_{1}^{(1)}$ and E$_{1}^{(2)}$ phonon modes; $G_{pi}$ (i=3,4) corresponds to the E$_{2}^{(1)}$ and E$_{2}^{(2)}$ phonon modes. The electronic Green's function $G_{ej}$ (j=1,2,3) has increasing energy in order; $G_{e1}$ and $G_{e3}$ are doublet-to-doublet CF excitations, and $G_{e2}$ is a doublet-to-singlet CF excitation. These Green's functions have a Lorentzian form: $-1/(\omega-\omega_0+i\gamma_0)$, in which $\omega_0$ is bare frequency and $\gamma_0$ is half width at half maximum (HWHM). The parameters $v_{ij}$ (i=1,2,3,4; j=1,2,3) represents the coupling strength between the phononic excitation $G_{pi}$ and electronic excitation $G_{ej}$.

We can describe the coupling in an alternative way to reduce the number of free parameters. We define the effective electric field $f_{i}$ (i=1,2,3,4) for each of the four phonon modes, and the effective electric moment $q_{jk}$ (j=1,2,3; k=1,2) for each of the three CF transitions. For $q_{jk}$, the first index j labels the three CF transitions; the second index k labels the symmetry of the corresponding electric moment: k=1 corresponds to E$_{1}$ symmetry and k=2 corresponds to E$_{2}$ symmetry. We then have $v_{ij}$=$f_{i}q_{j1}$ (i=1,2) and $v_{ij}$=$f_{i}q_{j2}$ (i=3,4). In this way, we reduce the number of needed coupling constants to ten from twelve.

The interacting Green's function $G_{E}$ for the E-symmetry components of the CF excitations is
\begin{equation}
G_{E}=(G_{E0}^{-1}-V_{E})^{-1}~.
\label{eq:GE}
\end{equation}
And the Raman responses in the XZ and RL scattering geometries are calculated in the following way:
\begin{equation}
\chi^{\prime\prime}_{XZ}\sim\Im T^{\dag}_{XZ}G_{E}T_{XZ}~,
\label{eq:ChiXZ}
\end{equation}
\begin{equation}
\chi^{\prime\prime}_{RL}\sim\Im T^{\dag}_{RL}G_{E}T_{RL}~.
\label{eq:ChiRL}
\end{equation}
In these two expressions, $T^{\dag}_{XZ}$ and $T^{\dag}_{RL}$ are the vertex of light scattering process in the XZ and RL scattering geometries, respectively:
\begin{equation}
T^{\dag}_{XZ}=\left(\begin{array}{ccccccc}t_{p1}&t_{p2}&0&0&t_{e1XZ}&t_{e2XZ}&t_{e3XZ}\end{array}\right)~,
\label{eq:TXZ}
\end{equation}
\begin{equation}
T^{\dag}_{RL}=\left(\begin{array}{ccccccc}0&0&t_{p3}&t_{p4}&t_{e1RL}&t_{e2RL}&t_{e3RL}\end{array}\right)~.
\label{eq:TRL}
\end{equation}

For the A$_{1}$/A-symmetry components of the CF excitations, the bare Green's function $G_{A0}$ and the coupling interaction $V_{A}$ are
\begin{equation}
G_{A0}=
\begin{pmatrix}
G_{p5} & 0      & 0      \\
0      & G_{e1} & 0      \\
0      & 0      & G_{e3} \\
\end{pmatrix}~,
\label{eq:GA0}
\end{equation}
\begin{equation}
V_{A}=
\begin{pmatrix}
0      & v_{51} & v_{53} \\
v_{51} & 0      & 0      \\
v_{53} & 0      & 0      \\
\end{pmatrix}~.
\label{eq:VA}
\end{equation}
Here the phononic Green's function $G_{p5}$ represents the A$_{1}$-symmetry phonon modes. Similar to the E-symmetry case, we have
\begin{equation}
G_{A}=(G_{A0}^{-1}-V_{A})^{-1}~,
\label{eq:GA}
\end{equation}
\begin{equation}
\chi^{\prime\prime}_{ZZ}\sim\Im T^{\dag}_{ZZ}G_{A}T_{ZZ}~,
\label{eq:ChiZZ}
\end{equation}
\begin{equation}
T^{\dag}_{ZZ}=\left(\begin{array}{ccc}t_{p5}&t_{e1ZZ}&t_{e3ZZ}\end{array}\right)~.
\label{eq:TZZ}
\end{equation}

In practice, we find that the XZ and ZZ spectra can be fitted without any coupling between CF excitations and phonon modes. This fact suggests that the coupling in E$_{1}$ and A$_{1}$ symmetry channels is weak, and in turn significantly simplifies the model because the fitting of the XZ and RL spectra are now decoupled. As the CF modes have the same energy and linewidth in different symmetry channels, we first fit the XZ spectrum and then use the obtained energy and linewidth for the fitting of RL and ZZ spectra.

For the XZ spectrum, we have
\begin{equation}
\chi^{\prime\prime}_{xz}\sim\Im T^{\dag}_{xz}G_{xz}T_{xz}~,
\label{eq:Chixz}
\end{equation}
in which
\begin{equation}
G_{xz}=
\begin{pmatrix}
G_{p1} & 0      & 0      & 0      & 0      \\
0      & G_{p2} & 0      & 0      & 0      \\
0      & 0      & G_{e1} & 0      & 0      \\
0      & 0      & 0      & G_{e2} & 0      \\
0      & 0      & 0      & 0      & G_{e3} \\ 
\end{pmatrix}~,
\label{eq:Gxz}
\end{equation}
and
\begin{equation}
T^{\dag}_{xz}=\left(\begin{array}{ccccc}t_{p1}&t_{p2}&t_{e1xz}&t_{e2xz}&t_{e3xz}\end{array}\right)~.
\label{eq:Txz}
\end{equation}

The RL spectrum is fitted with
\begin{equation}
\chi^{\prime\prime}_{rl}\sim\Im T^{\dag}_{rl}G_{rl}T_{rl}~,
\label{eq:Chirl}
\end{equation}
in which
\begin{equation}
G_{rl}=(G_{rl0}^{-1}-V_{rl})^{-1}~,
\label{eq:Grl}
\end{equation}
and 
\begin{equation}
T^{\dag}_{rl}=\left(\begin{array}{ccccc}t_{p3}&t_{p4}&t_{e1rl}&t_{e2rl}&t_{e3rl}\end{array}\right)~.
\label{eq:Trl}
\end{equation}
In Eq.~(\ref{eq:Grl}), $G_{rl0}$ and $V_{rl}$ have the following expression:
\begin{equation}
G_{rl0}=
\begin{pmatrix}
G_{p3} & 0      & 0      & 0      & 0      \\
0      & G_{p4} & 0      & 0      & 0      \\
0      & 0      & G_{e1} & 0      & 0      \\
0      & 0      & 0      & G_{e2} & 0      \\
0      & 0      & 0      & 0      & G_{e3} \\ 
\end{pmatrix}~,
\label{eq:Grl0}
\end{equation}
\begin{equation}
V_{rl}=
\begin{pmatrix}

& 0          & 0          & f_3*q_{12} & f_3*q_{22} & f_3*q_{32} \\
& 0          & 0          & f_4*q_{12} & f_4*q_{22} & f_4*q_{32} \\
& f_3*q_{12} & f_4*q_{12} & 0          & 0          & 0          \\
& f_3*q_{22} & f_4*q_{22} & 0          & 0          & 0          \\
& f_3*q_{32} & f_4*q_{32} & 0          & 0          & 0          \\
\end{pmatrix}~.
\label{eq:Vrl}
\end{equation}
In Eq.~(\ref{eq:Vrl}) we use the relationship $v_{ij}$=$f_{i}q_{j2}$ (i=3,4; j=1,2,3) to reduce the number of needed coupling constants to five from six.

For the ZZ spectrum, because the coupling is weak Eq.~(\ref{eq:ChiZZ}) is simplified to 
\begin{equation}
\chi^{\prime\prime}_{zz}\sim\Im T^{\dag}_{ZZ}G_{A0}T_{ZZ}~.
\label{eq:Chizz}
\end{equation}

\begin{figure}
\includegraphics[width=0.40\textwidth]{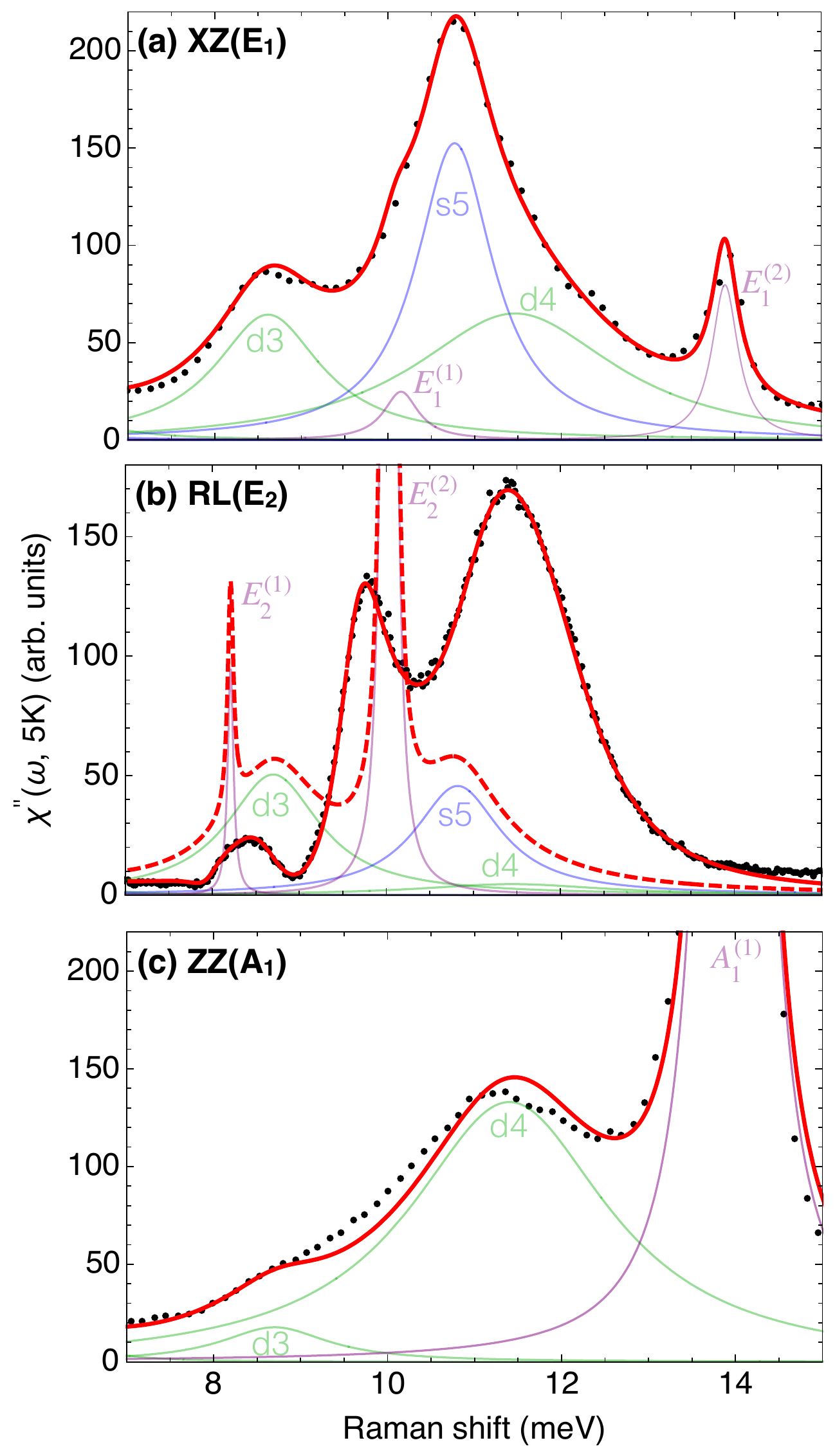}
\caption{\label{fig:CoupF}The fitting results for the coupled crystal-field (CF) and phonon modes. These spectra are the same as those shown in Fig.2 of main text. The XZ spectrum (a) is fitted with Eq.~(\ref{eq:Chixz}); the RL spectrum (b) with Eq.~(\ref{eq:Chirl}); the ZZ spectrum (c) with Eq.~(\ref{eq:Chizz}). The Raman data are represented by black dots. The red lines represent the fits. The blue lines show doublet-to-singlet CF oscillators labeled by the final singlet state; the green lines show doublet-to-doublet CF transitions; the phonon modes, labeled by their symmetry, are shown in purple. The dashed red line in panel (b) represents the sum of blue, green, and purple curves.}
\end{figure}

In Fig.~\ref{fig:CoupF} we compare the fits to the experimentally measured Raman data. The fitting curves match the data well. For RL spectrum [Fig.~\ref{fig:CoupF}(b)], the sum of the bare responses is rather different from the interacting response, which is significantly modulated by the couplings between pairs of phonon and CF modes. In appearance, the E$_{2}^{(1)}$ phonon mode and d3 CF mode "merge" into the vibronic feature at 8.5\,meV; the E$_{2}^{(2)}$ phonon mode and s5 CF mode exhibit level-repulsion behavior.

The fitting parameters can be found in Table~\ref{table:Para1}. Using the effective electric fields and effective electric moments, we can calculate the coupling strength in unit of energy between pairs of phonon and CF modes in RL spectrum (E$_{2}$ symmetry channel). Considering that the smaller the frequency difference between one pair of modes, the stronger the coupling effect is, we can construct dimensionless coupling constants by coupling strength divided by frequency difference of the same pair of modes. The relevant values are shown in Table~\ref{table:C}.

\begin{table}
\caption{\label{table:Para1}The values of the parameters obtained by fitting the XZ spectrum with Eq.~(\ref{eq:Chixz}), the RL spectrum with Eq.~(\ref{eq:Chirl}), and the ZZ spectrum with Eq.~(\ref{eq:ChiZZ}). Notice that although $f_{i}$ (i=3,4) and $q_{j2}$ (j=1,2,3) can have arbitrary units, $v_{ij}$=$f_{i}q_{j2}$ has the unit of meV.}
\begin{ruledtabular}
\begin{tabular}{ll}
Parameter (Unit)        & Value (Uncertainty)   \\
\hline
$t_{p1}$ (arb. units)   & 2.45(0.10)            \\
$\gamma_{p1}$ (meV)     & 0.242(0.015)          \\
$\omega_{p1}$ (meV)     & 10.15(0.01)           \\
\hline
$t_{p2}$ (arb. units)   & 3.83(0.01)            \\
$\gamma_{p2}$ (meV)     & 0.184(0.001)          \\
$\omega_{p2}$ (meV)     & 13.89(0.01)           \\
\hline
$t_{p3}$ (arb. units)   & 1.95(0.24)            \\
$\gamma_{p3}$ (meV)     & 0.041(0.017)          \\
$\omega_{p3}$ (meV)     & 8.21(0.02)            \\
$f_{3}$ (arb. units)    & -0.50(0.21)           \\
\hline
$t_{p4}$ (arb. units)   & 8.91(0.23)            \\
$\gamma_{p4}$ (meV)     & 0.031(0.022)          \\
$\omega_{p4}$ (meV)     & 10.03(0.02)           \\
$f_{4}$ (arb. units)    & -0.99(0.28)           \\
\hline
$t_{p5}$ (arb. units)   & 18.25(0.12)           \\
$\gamma_{p5}$ (meV)     & 0.223(0.003)          \\
$\omega_{p5}$ (meV)     & 13.98(0.02)           \\
\hline
$t_{e1xz}$ (arb. units) & 6.93(0.04)            \\
$t_{e1rl}$ (arb. units) & 5.73(0.03)            \\
$t_{e1zz}$ (arb. units) & 3.65(0.28)            \\
$\gamma_{e1}$ (meV)     & 0.703(0.006)          \\
$\omega_{e1}$ (meV)     & 8.65(0.02)            \\
$q_{12}$ (arb. units)   & 0.61(0.21)            \\
\hline
$t_{e2xz}$ (arb. units) & 9.10(0.17)            \\
$t_{e2rl}$ (arb. units) & 5.23(0.29)            \\
$\gamma_{e2}$ (meV)     & 0.58(0.04)            \\
$\omega_{e2}$ (meV)     & 10.80(0.11)           \\
$q_{22}$ (arb. units)   & 0.69(0.21)            \\
\hline
$t_{e3xz}$ (arb. units) & 9.85(0.17)            \\
$t_{e3rl}$ (arb. units) & 2.49(0.29)            \\
$t_{e3zz}$ (arb. units) & 10.54(0.08)           \\
$\gamma_{e3}$ (meV)     & 1.33(0.09)            \\
$\omega_{e3}$ (meV)     & 11.45(0.12)           \\
$q_{32}$ (arb. units)   & 0.92(0.25)            \\
\end{tabular}
\end{ruledtabular}
\end{table}

\section{The Specific-Heat Model\label{sec:ASH}}

We use four components to model the specific heat of TbInO$_3$ from 0 to 300\,K: (i) the nuclear contribution; (ii) the electronic contribution; (iii) the acoustic-phonon contribution; (iv) the optical-phonon contribution.

(i) The nuclear contribution is related to the transitions between the nuclear energy levels of Tb ions, and has the following form~\cite{Tari2003}:
\begin{multline}
C_N=\frac{R}{(k_BT)^2}\{\frac{1}{Z_N}\sum_{i}\epsilon_i^2e^{(-\epsilon_i/k_BT)}-\\
[\frac{1}{Z_N}\sum_{i}\epsilon_i e^{(-\epsilon_i/k_BT)}]^2\}~,
\label{eq:SHN}
\end{multline}
where R is the ideal gas constant; k$_B$ is the Boltzmann constant; T is temperature; $Z_N=\sum_{i}exp(-\epsilon_i/k_BT)$ is the partition function for the nuclear levels; $i$ (from 1 to 4) labels the individual nuclear energy levels; $\epsilon_i$ represents the energy of level $i$. The nuclear spin of $^{159}$Tb is I=3/2. Assuming that its four nuclear levels are evenly spaced~\cite{Heltemes1961,Yaouanc2011}, we find the splitting between each level is 0.003\,meV by fitting the specific heat below 0.5\,K.

(ii) The electronic contribution corresponds to the transitions between the CF states within the $^7F_6$ ground-state multiplet of Tb ions. The relevant formula has been discussed as Eq.~(\ref{eq:SHE}) in Sec.~\ref{sec:SH}.

(iii) We use Debye model for the acoustic-phonon contribution~\cite{Tari2003}. Because ferroelectric TbInO$_3$ has one A$_1$ and one E$_1$ acoustic modes, two Debye temperatures are used:
\begin{multline}
C_{D}=\frac{3R}{N}[(\frac{T}{\theta_{A_1}})^3\int_{0}^{\theta_{A_1}/T}\frac{x^4e^x}{(e^x-1)^2} dx+\\ 2(\frac{T}{\theta_{E_1}})^3\int_{0}^{\theta_{E_1}/T}\frac{x^4e^x}{(e^x-1)^2} dx]~,
\label{eq:SHD}
\end{multline}
in which N=6 is the number of chemical units per unit cell; $\theta_{A_1}$ is the Debye temperature for the A$_1$ mode and $\theta_{E_1}$ is that for the E$_1$ mode.

(iv) The optical-phonon contribution is accounted for by Einstein model~\cite{Tari2003}:
\begin{equation}
C_{Ein}=\frac{R}{N}\sum_{i}n_i(\frac{\hbar \omega_i}{k_BT})^2\frac{e^{\frac{\hbar \omega_i}{k_BT}}}{(e^{\frac{\hbar \omega}{k_BT}}-1)^2}~.
\label{eq:SHEin}
\end{equation}
In this formula, $i$ labels the individual optical phonon modes, and $n_i$ is the degeneracy of optical phonon mode $i$. Ferroelectric TbInO$_3$ has 38 Raman-active optical modes (9$A_{1}\oplus 14E_{1}\oplus 15E_{2}$) and 20 silent optical modes (5$A_{2}\oplus 10B_{1}\oplus 5B_{2}$). For the Raman-active optical modes, we use the experimentally determined energy listed in Table~\ref{table:P}; for the silent optical modes, we assume that they are evenly spaced from 10 to 80\,meV.

We subtract the nuclear, electronic, and optical-phonon contributions from the experimental specific-heat data, and then fit the subtracted data with Eq.~(\ref{eq:SHD}). We find $\theta_{A_1}$=180$\pm$50\,K and $\theta_{E_1}$=310$\pm$90\,K.

%\bibliography{Tb.bib}

\end{document}